\documentclass[letterpaper, preprint, paper,11pt]{AAS} 

\usepackage{bm}
\usepackage{amsmath}
\usepackage{amssymb}
\usepackage{subfigure}
\usepackage[colorlinks=true, pdfstartview=FitV, linkcolor=black, citecolor=black, urlcolor=black]{hyperref}
\usepackage{overcite}
\usepackage{footnpag}
\usepackage{subfigure}
\usepackage{comment}
\usepackage{graphicx}
\usepackage{booktabs}

\usepackage{tikz}
\usetikzlibrary{shapes.geometric, arrows.meta, positioning, fit, backgrounds, calc}

\definecolor{deepred}{RGB}{210, 0, 0} 

\PaperNumber{AAS-26-909}

\begin{document}

\title{ANALYTICAL CONFIDENCE BOUNDARIES FOR NON-GAUSSIAN UNCERTAINTY IN PERTURBED SPACECRAFT DYNAMICS}

\author{Niccolò Michelotti\thanks{PhD Student, Department of Aerospace Science and Technology, Politecnico di Milano, via La Masa, 34, Milan, 20156, Italy. niccolo.michelotti@polimi.it}, Ethan R. Burnett\thanks{Assistant Professor, Department of Aerospace Engineering and Engineering Mechanics, University of Texas at Austin, 2617 Wichita St., Austin TX, 78712, USA. ethan.burnett@utexas.edu. }, and Francesco Topputo\thanks{Full Professor, Department of Aerospace Science and Technology, Politecnico di Milano, via La Masa, 34, Milan, 20156, Italy. francesco.topputo@polimi.it. AIAA Senior Member.} }
\maketitle{}

\begin{abstract}
This work investigates nonlinear uncertainty propagation in perturbed astrodynamics, focusing on the rapid characterization of non-Gaussian distributions and the construction of three-dimensional ``banana-shaped'' confidence boundaries. To bridge the gap between computationally intensive high-fidelity methods and inaccurate linear approximations, this paper introduces a fully analytical, sample-free framework for higher-order moments extraction. Leveraging Differential Algebra to bypass repeated numerical integration, statistical moments are extracted analytically via Isserlis' theorem and a monomial-to-Hermite basis transformation. A pair-product projection strategy is exploited to overcome the severe computational bottleneck of full fourth-order tensor contractions and compute only relevant terms via efficient polynomial algebra. The extracted skewness and kurtosis components directly parameterize non-elliptical confidence geometries that capture spatial bending and out-of-plane coupling of typical non-Gaussian distributions in astrodynamics. The approach is validated in high-fidelity environments including a cislunar Near-Rectilinear Halo Orbit and close-proximity trajectories around Apophis during Earth's flyby, where the analytical approach achieves geometric accuracy comparable to expensive Monte Carlo simulations while reducing computational runtime by orders of magnitude.
\end{abstract}

\section{Introduction}
\label{sec:introduction}
Reliable and efficient uncertainty quantification (UQ) is a persistent challenge in modern astrodynamics. As mission architectures increasingly target highly nonlinear and weakly stable regimes, capturing the true evolution of state uncertainty becomes critical for the robustness and efficiency of the mission. In cislunar environments, such as Near-Rectilinear Halo Orbits (NRHOs), spacecraft are subjected to sensitive multi-body gravitational fields and repeated low-altitude perilune passages \cite{wolf_multi-fidelity_2022}. Similarly, close-proximity operations around small bodies, such as the asteroid Apophis, are dominated by irregular gravity fields and rapid dynamical timescales. In these environments, initial Gaussian state uncertainties quickly deform into non-Gaussian distributions driven by strong localized gravitational gradients and resonances \cite{scheeres2016orbital}. Consequently, standard geometric descriptions based uniquely on the mean and covariance fail to capture relevant features of the distribution. Accurately bounding the spacecraft's state is a fundamental requirement for robust trajectory optimization and space domain awareness. In the presence of non-Gaussian distributions, this requires the evaluation of higher-order statistical moments, such as skewness and kurtosis, to properly represent the asymmetry and tail-stretching of the probability density function.

Historically, Monte Carlo (MC) methods have served as the highest-fidelity benchmark for capturing these non-Gaussian features. However, their reliance on thousands of numerical integrations makes them computationally expensive for rapid and preliminary analyses, large-scale trajectory design campaigns, or onboard applications for autonomous spacecraft. Conversely, linear covariance (LinCov) propagation is computationally efficient but fundamentally misrepresents uncertainty distribution in highly nonlinear flows, often underestimating the actual confidence boundaries \cite{michelotti_uncertainty_2024}. To balance accuracy and computational cost, higher-fidelity nonlinear propagation techniques have been investigated thoroughly \cite{luo2017review}. Advanced techniques such as Conjugate Unscented Transform (CUT) \cite{adurthi2012conjugate}, Polynomial Chaos Expansion (PCE) \cite{jones2013nonlinear} and Gaussian Mixture Models (GMMs) \cite{vittaldev2016spacecraft} have been implemented in several applications for astrodynamics, orbital determination and conjunction assessment.  Nevertheless, all these methods still require repeated numerical evaluations of the full dynamics to compute local expansions. For repetitive UQ queries, this repeated sampling remains a relevant computational limitation.

To circumvent the need for repeated numerical integration, Differential Algebra (DA) \cite{valli_nonlinear_2013} and State Transition Tensors (STTs) \cite{boone2023directional} have emerged as effective tools for orbital uncertainty propagation. By computing high-order Taylor expansions of the dynamical flow map, the complex numerical integration is replaced by the evaluation of a polynomial map. When combined with mathematical identities such as Isserlis' theorem (or Wick's probability theorem) \cite{isserlis1918formula}, DA enables the sample-free extraction of statistical moments directly from the expansion coefficients \cite{acciarini_nonlinear_2025}. However, the extraction of high-order moments inherently demands the construction and contraction of dense high-dimensional tensors. This combinatorial growth becomes particularly relevant starting from the fourth-order moment (kurtosis), introducing an $\mathcal{O}(N^4)$ computational and memory cost, limiting the scalability of the method for higher-order spatial reconstructions in full 6-degree-of-freedom state spaces.

This work proposes a fully analytical sample-free optimized framework for non-Gaussian uncertainty approximation and three-dimensional confidence boundary reconstruction. First, the paper implements and compares sample-free methods for extracting higher-order moments from DA maps, highlighting the specific sparsity advantages of transforming the standard monomial basis into an orthogonal probabilist's Hermite basis. Second, to address the $\mathcal{O}(N^4)$ tensor limitation during kurtosis extraction, a ``project-then-product'' strategy is introduced. By projecting the maps onto principal bending axes and exploiting the orthogonality of the Polynomial Chaos basis, the computational complexity is reduced to one-dimensional polynomial algebra. Finally, the analytically derived skewness and kurtosis are utilized to reconstruct non-elliptical, three-dimensional confidence geometries, extending a previous derivation only provided in two-dimensional slices of the problem variables\cite{BurnettBooneCDC26}. This ``banana'' approximation captures the complex spatial bending and out-of-plane coupling induced by highly perturbed astrodynamical environments. This or analogous analytical methods are best used for non-Gaussian distributions where linear covariance methods are no longer valid, but the distribution remains unimodal and easily parameterizable. Whereas this paper focuses on the efficient real-time computation of higher moments for a non-Gaussian confidence contour, Ref.~\citenum{AAS26a}, presented at the same conference, focuses on stochastic guidance applications of the analytic non-Gaussian confidence contour concept. However, that work does not employ differential algebra nor the sample-free methods developed here.

The remainder of this paper is organized as follows. Section \ref{sec:confidence_boundary} outlines the formulation of the 3D banana-shaped confidence boundary reconstruction. Section \ref{sec:analytical_higher_order} details the mathematical framework for analytical higher-order moment extraction, including the DA-Isserlis formulation, the Hermite basis transformation, and the pair-product strategy. Section \ref{sec:numerical_results} presents numerical results validating the proposed framework in two representative environments: a cislunar NRHO and close-proximity trajectories around the asteroid Apophis during its upcoming Earth flyby. Results highlight the trade-offs between propagation accuracy and computational runtime of the proposed approach with the performance of state-of-the-art methods. Finally, Section \ref{sec:conclusion} concludes the paper.

\section{Confidence Boundary Reconstruction}
\label{sec:confidence_boundary}
To motivate the development of the analytical moment-extraction framework presented in Section \ref{sec:analytical_higher_order}, we first outline the target geometric application: the analytical boundary reconstruction.
This formulation leverages the higher-order statistical moments (skewness and kurtosis) to deform the principal axes of the local covariance ellipsoid, explicitly capturing the non-Gaussian stretching and folding of the propagated uncertainty. The foundational formulation of this theory was presented by Ref.~\citenum{BurnettBooneCDC26} to approximate two-dimensional ``banana-shaped'' distributions commonly found in astrodynamics. This section details the reconstruction process and formalizes its extension to three-dimensional spatial volumes to capture out-of-plane nonlinear couplings. 

Once the covariance and higher-order moments are computed, the 3D boundary reconstruction proceeds through the following steps:

\begin{enumerate}
    \item \textbf{Eigendecomposition and Axis Sorting:}\\
    To apply the geometric deformations correctly, a local orthogonal frame must be established along the principal axes of the uncertainty distribution. This is achieved via the eigenvalue decomposition of the $3 \times 3$ spatial covariance matrix, $\bm{P} = \bm{V} \bm{\Sigma} \bm{V}^\top$. The eigenvectors are sorted based on their associated variance and higher-order couplings. The eigenvector corresponding to the largest eigenvalue (maximum variance) is defined as the primary ``length'' direction. The secondary ``width'' axis is selected as the eigenvector exhibiting the strongest skewness coupling (cross-skewness) with the length direction, while the remaining orthogonal eigenvector corresponds to the ``thickness'' direction. Let $\bm{u}$, $\bm{v}$, and $\bm{w}$ denote the local whitened basis vectors corresponding to these length, width, and thickness directions, respectively. 

    \item \textbf{Moment Projection:}\\
    The geometric deformation of the boundary is governed by the projection of the skewness and kurtosis tensors onto this local frame. The relevant coupling quantities are mathematically defined as the tensor projections:
    \begin{subequations}
    \begin{align}
        E_{uuu}  &:= E[\bm{u}^3] = \sum_{i,j,k} u_i u_j u_k \, \mathcal{M}^{(3)}_{ijk} \\
        E_{uuv}  &:= E[\bm{u}^2\bm{v}] = \sum_{i,j,k} u_i u_j v_k \, \mathcal{M}^{(3)}_{ijk} \\
        E_{uuw}  &:= E[\bm{u}^2\bm{w}] =  \sum_{i,j,k} u_i u_j w_k \, \mathcal{M}^{(3)}_{ijk} \\
        E_{uuuu} &:=  E[\bm{u}^4] = \sum_{i,j,k,l} u_i u_j u_k u_l \, \mathcal{M}^{(4)}_{ijkl}
    \end{align}
    \label{eq:relevant_terms}
    \end{subequations}
    where $E_{uuu}$ captures the lengthwise asymmetry, $E_{uuv}$ and $E_{uuw}$ capture the in-plane and out-of-plane bending, and $E_{uuuu}$ quantifies the lengthwise elongation. 
    In standard practice, evaluating these specific projections requires the costly construction of the full dense tensors $\mathcal{M}^{(3)}$ and $\mathcal{M}^{(4)}$. Bypassing this computational bottleneck is the primary focus of the mathematical framework developed in Section \ref{sec:analytical_higher_order}, which computes these exact quantities directly via a highly efficient 1D "project-then-product" strategy.

    \item \textbf{3D Shape Corrections:}\\
    In the local whitened frame, the undeformed confidence boundary is initialized as a sphere parameterized by standard spherical coordinates:
    \begin{equation}
        u = k\cos\theta\sin\phi, \quad v = k\sin\theta\sin\phi, \quad w = k\cos\phi
        \label{eq:spherical_coord}
    \end{equation}
    where $k$ is the desired confidence level parameter (e.g., $k=3$ for a $3\sigma$ equivalent boundary). 
    
    The first correction applies an in-plane bending deformation, driven by the cross-skewness $E_{uuv}$ (coupling length and width) and moderated by the kurtosis $E_{uuuu}$ (lengthwise elongation). This is modeled as a quadratic deformation along the width direction:
    \begin{equation}
        \Delta v_b = \alpha (u^2 - 1), \qquad \alpha = \frac{E_{uuv}}{E_{uuuu} - 1}
    \end{equation}
    
    The second correction introduces an asymmetry along the primary length direction to account for the heavy-tailed nature of the distribution. This shift is derived from the Cornish-Fisher expansion \cite{cornish1938moments}:
    \begin{equation}
        \Delta u_{CF} = c_k \frac{u^2}{k^2}, \qquad c_k = \frac{1}{6} E_{uuu} (k^2 - 1)
    \end{equation}
    
    The third correction constitutes the extension to the three-dimensional case. Highly perturbed astrodynamical environments often induce spatial torsion, requiring an out-of-plane bending correction driven by $E_{uuw}$, which captures the coupling between the length and thickness directions. It is computed analogously to the in-plane deformation:
    \begin{equation}
        \Delta w_{b} = \beta (u^2 - 1), \qquad \beta = \frac{E_{uuw}}{E_{uuuu} - 1}
    \end{equation}
    
    By superimposing these corrections onto the base sphere (Eq.\ \eqref{eq:spherical_coord}), the deformed local coordinates $(u', v', w')$ are obtained:
    \begin{equation}
        u' = u + \Delta u_{CF}, \quad v' = v + \Delta v_{b}, \quad w' = w + \Delta w_{b}
    \end{equation}

    \item \textbf{Global Frame Transformation:}\\
    Finally, the boundary is transformed from the local standard normal space back into the physical state space. The deformed coordinates are scaled by the principal standard deviations ($\bm{\Sigma}^{1/2}$), rotated back to the global frame using the sorted eigenvector matrix $\bm{V}$, and translated by the propagated mean vector $\bm{\mu}$:
    \begin{equation}
        \bm{x}_{bound} = \bm{\mu} + \bm{V} \bm{\Sigma}^{1/2} \begin{bmatrix} u' \\ v' \\ w' \end{bmatrix}
    \end{equation}
\end{enumerate}

The outlined approach yields a continuous, analytical 3D confidence boundary that reliably encompasses the true uncertainty geometry in nonlinear regimes, circumventing the limitations of standard ellipsoidal approximations. Furthermore, this geometric framework is inherently modular. While the present implementation relies on quadratic bending and first-order Cornish-Fisher expansions, the formulation can theoretically be extended to incorporate alternative analytical deformation models. By superimposing additional shape corrections directly into the local coordinate frame, more complex topological features could be captured without altering the underlying moment-extraction architecture.

\section{Analytical Higher-Order Moment Extraction}
\label{sec:analytical_higher_order}
This section details the analytical framework developed to estimate the higher-order moments of propagated uncertainty distributions. The proposed approaches bypass any sampling step and extract statistical moments directly from the algebraic structure of the nonlinear flow maps.

Transitioning from sample-based evaluations to purely analytical formulations offers advantages for both uncertainty quantification and trajectory design. Traditional sampling techniques are affected by statistical noise, slow $\mathcal{O}(1/\sqrt{N})$ convergence rates of MC simulations, and the introduction of non-smooth stochastic gradients. These artifacts are highly problematic for gradient-based robust optimization solvers, which rely on smooth, continuous design spaces to satisfy exact optimality conditions. Furthermore, deterministic moment extraction guarantees exact reproducibility and facilitates the formulation of rigorous chance constraints for safety-critical spaceflight operations \cite{AAS26a}. 

To achieve these benefits without incurring the severe memory overhead typically associated with dense multidimensional tensors, three analytical formulations of increasing computational efficiency are presented below.  First, the fundamental combination of Differential Algebra (DA) techniques and Isserlis' theorem is presented. Next, a Monomial-to-Hermite basis transformation is introduced to increase the sparsity of the expected value tensors. Finally, a ``project-then-product'' strategy is derived, enabling the direct computation of only the strictly relevant terms of the higher-order moment tensors.

\subsection{Differential Algebra Maps and Isserlis' Theorem}
\label{sec:da_isserlis}
Consider a general nonlinear dynamical system governed by the ordinary differential equation:\begin{equation}\dot{\bm{x}} = \bm{f}(\bm{x}, t)\end{equation}where $\bm{x}$ is the state vector. Given a nominal initial state $\bm{x}_0$ and an initial deviation $\delta \bm{x}_0$, the state at a future time $t$ is defined by the flow map:\begin{equation}\bm{x}(t) = \bm{\phi}(t; \bm{x}_0 + \delta \bm{x}_0, t_0)
\end{equation}
In standard Monte Carlo methods, this equation requires repeated numerical integration for each sampled initial condition. Differential Algebra provides a mathematical framework to bypass this limitation, substituting integrations of thousands of samples with inexpensive Taylor polynomial evaluations. Initializing the state as a DA variable yields a high-order Taylor expansion of the flow map with respect to the initial deviations:
\begin{equation}
    \delta \bm{x}(t) = \sum_{|\bm{p}| \le k} \bm{M}_{\bm{p}}(t) \delta \bm{x}_0^{\bm{p}}
\end{equation}
where $\bm{M}_{\bm{p}}(t)$ are the deterministic Taylor coefficients up to order $k$, and $\bm{p}$ is a multi-index representing the variable exponents. To evaluate the statistical moments of the propagated state, let the initial uncertainty be described by a Gaussian random vector $\delta \bm{x}_0 \sim \mathcal{N}(\bm{0}, \bm{P}_0)$. If this correlated deviation is utilized directly, computing higher-order moments requires the general form of Isserlis' theorem (or Wick's probability theorem) \cite{isserlis1918formula}. The general theorem states that the expected value of the product of zero-mean correlated Gaussian variables is the sum over all possible distinct ways of partitioning the indices into unordered pairs of their covariances \cite{acciarini_nonlinear_2025}:

\begin{equation}
    \mathbb{E}[\delta x_{d_1} \dots \delta x_{d_k}] = \begin{cases}
        0 & \text{if } k \text{ is odd} \\
        \sum_{p \in \mathcal{P}} \prod_{\{u,v\} \in p} (\bm{P}_0)_{d_u d_v} & \text{if } k \text{ is even}
    \end{cases}
    \label{eq:isserlis_general}
\end{equation}
To simplify the combinatorial evaluation of Eq.\ \eqref{eq:isserlis_general}, a whitening process is performed. The initial deviations are mathematically decorrelated prior to the DA expansion using the Cholesky decomposition $\bm{P}_0 = \bm{L}\bm{L}^T$. This defines a set of independent standard normal variables $\bm{\xi} \sim \mathcal{N}(\bm{0}, \bm{I})$, such that $\delta \bm{x} = \bm{L}\bm{\xi}$. The DA engine then maps this linear transformation through the dynamics to yield a high-order Taylor expansion of the final state component $x_i$ entirely within the whitened standard normal space:
\begin{equation}
    x_i(\bm{\xi}) = \sum_{|\bm{p}| \le k} c_{i,\bm{p}} \bm{\xi}^{\bm{p}}
    \label{eq:da_expansion}
\end{equation}
where $c_{i,\bm{p}}$ are the deterministic map coefficients evaluated via DA, $\bm{p} \in \mathbb{N}^m$ is a multi-index vector of exponents, and $\bm{\xi}^{\bm{p}} = \prod_{d=1}^{m} \xi_d^{p_d}$ forms the standard monomial basis.

Because the variables in $\bm{\xi}$ are independent standard normals, their covariance matrix is the identity matrix $\bm{I}$, meaning $\mathbb{E}[\xi_i \xi_j] = \delta_{ij}$. When Isserlis' theorem is applied in this whitened space, any combinatorial pairing in the summation that links two distinct dimensions ($i \neq j$) evaluates to zero and vanishes. The only surviving terms are those that pair a variable strictly with itself ($\mathbb{E}[\xi_i^2] = 1$). For a univariate term $\xi_d^n$, the number of unique pairings of $n$ identical items is given by the double factorial, resulting in a straightforward 1D expectation:
\begin{equation}
    \mathbb{E}[\xi_d^n] = \begin{cases}
        0 & \text{if } n \text{ is odd} \\
        (n-1)!! & \text{if } n \text{ is even}
    \end{cases}
    \label{eq:isserlis_1d}
\end{equation}
Furthermore, due to the independence of the components, the expectation of any multidimensional monomial basis term reduces to the product of its univariate expectations:
\begin{equation}
    \mathbb{E}[\bm{\xi}^{\bm{p}}] = \mathbb{E}\left[\prod_{d=1}^m \xi_d^{p_d}\right] = \prod_{d=1}^m \mathbb{E}[\xi_d^{p_d}]
    \label{eq:isserlis_multivariate}
\end{equation}
By applying the linearity of the expectation operator to the DA expansion in Eq.\ \eqref{eq:da_expansion}, the raw statistical moments of the propagated state can be evaluated analytically. The first two raw moments are obtained as:
\begin{subequations}
\begin{align}
    \mu_i &= \mathbb{E}[x_i(\bm{\xi})] = \sum_{|\bm{p}| \le k} c_{i,\bm{p}} \, \mathbb{E}[\bm{\xi}^{\bm{p}}] \\
    \mathcal{R}^{(2)}_{ij} &= \mathbb{E}[x_i(\bm{\xi}) x_j(\bm{\xi})] = \sum_{|\bm{p}| \le k} \sum_{|\bm{q}| \le k} c_{i,\bm{p}} \, c_{j,\bm{q}} \, \mathbb{E}[\bm{\xi}^{\bm{p}+\bm{q}}]
\end{align}
\label{eq:mean_covariance}
\end{subequations}
where the combined expectation term $\mathbb{E}[\bm{\xi}^{\bm{p}+\bm{q}}]$ is evaluated using the double factorial property defined in Eqs.\ \eqref{eq:isserlis_1d} and \eqref{eq:isserlis_multivariate}. The central covariance matrix is then recovered as $\bm{P}_{ij} = \mathcal{R}^{(2)}_{ij} - \mu_i \mu_j$.

From Eq.\ \eqref{eq:mean_covariance}, it is possible to generalize the moment extraction process. The evaluation of any $n$\textsuperscript{th} order raw moment requires an $n$-fold summation over the product of $n$ map coefficients, weighted by the expectation of the combined monomial exponents. To formalize this, a generic $n$\textsuperscript{th} order \textit{weight tensor} $\mathcal{W}^{(n)}$ is defined to encode the statistical expectations of the combined monomial basis terms:
\begin{equation}
    \mathcal{W}^{(n)}_{\bm{p}_1 \dots \bm{p}_n} = \mathbb{E}[\bm{\xi}^{\bm{p}_1 + \dots + \bm{p}_n}]
\end{equation}
Because these weight tensors depend exclusively on the dimensionality $m$ and the expansion order $k$, they are independent of the orbital dynamics and can be computed offline. Consequently, the evaluation of the third and fourth raw moments ($\mathcal{R}^{(3)}$ and $\mathcal{R}^{(4)}$) becomes a set of tensor contractions between the DA coefficients $\bm{c}$ and the precomputed weight tensors:
\begin{equation}
    \mathcal{R}^{(3)}_{abc} = \sum_{\bm{p},\bm{q},\bm{r}} c_{a,\bm{p}} \, c_{b,\bm{q}} \, c_{c,\bm{r}} \, \mathcal{W}^{(3)}_{\bm{p}\bm{q}\bm{r}}, \qquad 
    \mathcal{R}^{(4)}_{abcd} = \sum_{\bm{p},\bm{q},\bm{r},\bm{s}} c_{a,\bm{p}} \, c_{b,\bm{q}} \, c_{c,\bm{r}} \, c_{d,\bm{s}} \, \mathcal{W}^{(4)}_{\bm{p}\bm{q}\bm{r}\bm{s}}
\label{eq:tensor_contractions}
\end{equation}
Finally, the required central moments of skewness ($\mathcal{M}^{(3)}$) and kurtosis ($\mathcal{M}^{(4)}$) are recovered from these raw multidimensional tensors via standard binomial expansions centered around the propagated mean \cite{acciarini_nonlinear_2025}.

\subsection{Monomial-To-Hermite Basis Transformation}
\label{sec:monomial_to_hermite}
While the Isserlis evaluation of the monomial basis is exact, computational efficiency can be improved by transforming the DA map into an orthogonal probabilist's Hermite polynomial basis. This derives an analytical Polynomial Chaos Expansion (PCE) directly from the DA map, bypassing the need for sampling or regression techniques \cite{sudret2008global}.

Because the map in Eq.\ \eqref{eq:da_expansion} is already expressed in terms of independent standard normal variables $\bm{\xi}$, any 1D monomial $\xi_d^n$ can be mapped exactly to a linear combination of Hermite polynomials $He_k(\xi_d)$ of the same parity \cite{olver2010nist}:
\begin{equation}
    \xi_d^n = \sum_{k \in \{n, n-2, \dots\}} w_{n,k} He_k(\xi_d)
\end{equation}
where the projection weights $w_{n,k}$ are given by:
\begin{equation}
    w_{n,k} = \frac{n!}{k! \, 2^{\frac{n-k}{2}} \, (\frac{n-k}{2})!}
\end{equation}
Applying this mapping to the multivariate case, the output map is rewritten as an analytical Hermite PCE:
\begin{equation}
    x_i(\bm{\xi}) = \sum_{\bm{\alpha}} c_{i,\bm{\alpha}}^{\text{PCE}} \Psi_{\bm{\alpha}}(\bm{\xi})
\end{equation}
where $\Psi_{\bm{\alpha}}(\bm{\xi}) = \prod_{d=1}^m He_{\alpha_d}(\xi_d)$ are the multivariate Hermite polynomials, and $c_{i,\bm{\alpha}}^{\text{PCE}}$ are the analytically transformed coefficients.

A key property of this basis is its orthogonality with respect to the standard normal Gaussian measure. The squared norm of a multivariate basis polynomial $\Psi_{\bm{\alpha}}$ evaluates directly to the factorial of its multi-index:
\begin{equation}
    \langle \Psi_{\bm{\alpha}}, \Psi_{\bm{\alpha}} \rangle = \mathbb{E}[\Psi_{\bm{\alpha}}^2] = \bm{\alpha}! = \prod_{d=1}^m \alpha_d!
\end{equation}

This PCE framework is effective for moment extraction because it leverages this basis orthogonality. For instance, the expectation of the triple product of 1D Hermite polynomials evaluates to exactly zero unless strict degree parity and triangle-inequality conditions are met \cite{azor1982combinatorial}:
\begin{equation}
    \mathbb{E}[He_u(\xi) He_v(\xi) He_w(\xi)] = \begin{cases} 
      0 & \text{if } u+v+w \text{ is odd, or } s < \max(u,v,w) \\
      \frac{u! v! w!}{(s-u)! (s-v)! (s-w)!} & \text{otherwise}
   \end{cases} 
\end{equation}
where $s = \frac{1}{2}(u+v+w)$.  
Evaluating higher-order moments involves substituting the PCE coefficients $c^{\text{PCE}}$ and the corresponding Hermite weight tensors $\mathcal{W}_{\text{PCE}}$ into the generalized contraction equations (Eq.\ \eqref{eq:tensor_contractions}). 

While the triple product efficiently populates the third-order tensor $\mathcal{W}^{(3)}_{\text{PCE}}$ for skewness, evaluating kurtosis requires the expectation of quadruple products to populate $\mathcal{W}^{(4)}_{\text{PCE}}$. Although analytical formulas exist for quadruple products, the combinatorial size of the resulting full tensor remains computationally demanding in high-dimensional spaces. Nevertheless, the Hermite weight tensors are significantly sparser than their monomial counterparts. For a 6D state and a 3rd-order DA expansion, the fill factor of the weight tensors (i.e., the percentage of non-zero elements) is approximately 0.6\% in the Hermite basis, compared to 2.5\% in the monomial basis. This sparsity reduces the computational cost of the tensor contractions, but the explicit construction of the full fourth-order tensor motivates further optimization, as addressed in the following section.

\subsection{Pair-product Strategy for Projected Moments}
\label{sec:pair-product}
Generalized tensor contractions provide a complete statistical characterization of the propagated state; however, constructing the full fourth-order moment tensor $\mathcal{W}^{(4)}$ for a 6D state space incurs a massive $\mathcal{O}(N^4)$ computational and memory cost. When the objective is strictly to characterize geometric features of the uncertainty distribution, such as extracting moments along the principal bending directions for boundary reconstruction, evaluating the entire dense tensor is mathematically redundant. To circumvent this limitation, a novel ``project-then-product'' strategy is introduced. Previous works, such as the efficient polynomial expansion strategies by Lefebvre (2021) \cite{lefebvre2021moment}, have demonstrated how to algebraically reduce the cost of computing high-order moments for scalar outputs by exploiting the orthogonality of the expansion basis. However, extending these scalar methods to evaluate the high-order cross-moments of a multi-dimensional state vector still triggers a combinatorial explosion of the cross-terms. Instead of computing these multi-dimensional cross-moments directly, this approach leverages geometric dimensionality reduction. To isolate specific non-Gaussian characteristics, first the multi-dimensional PCE polynomials are projected onto these specific 1D principal directions. Then, the recursive polynomial product identity are applied strictly to this 1D projection \cite{lefebvre2021moment}. By prioritizing the geometric projection, this strategy completely circumvents the full 6D tensor contraction, reducing a highly complex multi-dimensional moment problem to a lightweight 1D polynomial algebra operation.

Let $\bm{u}$, $\bm{v}$, and $\bm{w}$ be the selected principal directions, defined by the eigenvectors corresponding to the length, width, and thickness of the spatial covariance distribution. Due to the linearity of the projection operator, the full multidimensional expansion can be mapped directly into 1D, mean-centered scalar polynomial maps:
\begin{equation}
    z_u(\bm{\xi}) = \bm{u}^\top(\bm{x}(\bm{\xi}) - \bm{\mu}) = \sum_{|\bm{i}| \le k} \gamma_{u,\bm{i}} \Psi_{\bm{i}}(\bm{\xi}) 
    \label{eq:expansion_projection}
\end{equation}
where $\gamma_{u,\bm{i}}$ are the projected coefficients obtained by taking the dot product of the analytical Hermite PCE coefficients $\bm{c}^{\text{PCE}}_{\bm{i}}$ with the principal direction $\bm{u}$. Because the map is mean-centered, the zeroth-degree coefficient $\gamma_{u,\bm{0}}$ evaluates strictly to zero. Analogous definitions to Eq.\ \eqref{eq:expansion_projection} with respect to the other principal directions $\bm{v}$ and $\bm{w}$, yield the projected coefficient vectors $\bm{\gamma}_v$ and $\bm{\gamma}_w$ and the scalar maps $z_v$ and $z_w$.

To calculate the specific higher-order projected moments required for the confidence boundary deformation (e.g., $E_{uuu}, E_{uuv}, E_{uuw}, E_{uuuu}$), a standard evaluation would still require the contraction of the full third- and fourth-order weight tensors. Rather than performing these computationally expensive high-order contractions, complex statistical moments can be algebraically rewritten as the expected value of lower-order polynomials. 
Following the recursive decomposition strategy \cite{lefebvre2021moment}, this is achieved by defining an intermediate \textit{pair-product polynomial}. For instance, the squared deviation along the primary bending axis $\bm{u}$ is defined as:
\begin{equation}
    p_{uu}(\bm{\xi}) = z_u(\bm{\xi}) z_u(\bm{\xi})
    \label{eq:double_product}
\end{equation}
Because the Hermite polynomials form a complete orthogonal basis, this squared polynomial can be expressed as its own expansion, $p_{uu}(\bm{\xi}) = \sum c_{p_{uu}, \bm{k}} \Psi_{\bm{k}}(\bm{\xi})$. The unknown coefficients $\bm{c}_{p_{uu}}$ are found by multiplying both sides of Eq.\ \eqref{eq:double_product} by the $\bm{k}$-th basis function and taking the expected value. Substituting the definition of $z_u$ and applying the linearity of the expectation operator yields:
\begin{equation}
    \mathbb{E}[p_{uu} \Psi_{\bm{k}}] = \mathbb{E}[z_u z_u \Psi_{\bm{k}}] = \sum_{\bm{i},\bm{j}} \gamma_{u,\bm{i}} \gamma_{u,\bm{j}} \mathbb{E}[\Psi_{\bm{i}} \Psi_{\bm{j}} \Psi_{\bm{k}}] 
\end{equation}
Recognizing that $\mathbb{E}[\Psi_{\bm{i}} \Psi_{\bm{j}} \Psi_{\bm{k}}]$ matches the analytical definition of the third-order weight tensor $\mathcal{W}^{(3)}_{\bm{i}\bm{j}\bm{k}}$, the coefficients of the pair-product polynomial are recovered analytically:
\begin{equation}
    c_{p_{uu}, \bm{k}} = \frac{1}{\mathbb{E}[\Psi_{\bm{k}}^2]} \sum_{\bm{i},\bm{j}} \gamma_{u,\bm{i}} \gamma_{u,\bm{j}} \mathcal{W}^{(3)}_{\bm{i}\bm{j}\bm{k}} 
\end{equation}
where $\mathbb{E}[\Psi_{\bm{k}}^2] = \bm{k}!$ is the squared norm of the multi-dimensional Hermite basis term. This formulation demonstrates that the squared polynomial $p_{uu}$ can be assembled explicitly using only the precomputed third-order tensor. 

The primary computational advantage of this projection strategy lies in the extensive reuse of the 1D $p_{uu}$ polynomial. Once the coefficients $\bm{c}_{p_{uu}}$ are evaluated, the extraction of all primary and cross-coupling moments required for the boundary reconstruction reduces to computing the expected value of their product. Because $\Psi_{\bm{k}}$ is an orthogonal basis, $\mathbb{E}[\Psi_{\bm{i}} \Psi_{\bm{k}}] = 0$ for $\bm{i} \neq \bm{k}$, which naturally zeroes out all cross-terms and yields:
\begin{subequations}
    \begin{align}
    E_{uuu} &= \mathbb{E}[p_{uu} z_u] = \sum_{\bm{k}} c_{p_{uu}, \bm{k}} \, \gamma_{u,\bm{k}} \, \bm{k}! \\
    E_{uuv} &= \mathbb{E}[p_{uu} z_v] = \sum_{\bm{k}} c_{p_{uu}, \bm{k}} \, \gamma_{v,\bm{k}} \, \bm{k}! \\
    E_{uuw} &= \mathbb{E}[p_{uu} z_w] = \sum_{\bm{k}} c_{p_{uu}, \bm{k}} \, \gamma_{w,\bm{k}} \, \bm{k}! \\
    E_{uuuu} &= \mathbb{E}[p_{uu} p_{uu}] = \sum_{\bm{k}} c_{p_{uu}, \bm{k}}^2 \, \bm{k}! 
\end{align}
\label{eq:pair-product_relevant_terms}
\end{subequations}

Because $p_{uu}$ naturally encodes the nonlinear variance distribution along the primary bending axis $u$, the cross-coupling terms ($E_{uuv}$ and $E_{uuw}$) are retrieved directly by multiplying the $p_{uu}$ coefficients with the 1D coefficients of the secondary and tertiary axes ($z_v$ and $z_w$). By projecting the maps before calculating the expectations, this methodology bypasses the evaluation of the dense fourth-order tensor $\mathcal{W}^{(4)}$ entirely. 

It is necessary to address a theoretical discrepancy introduced by the pair-product approach. Mathematically, squaring a projected polynomial map of degree $k$ yields a polynomial of degree $2k$. However, because the pair-product coefficients $\bm{c}_{p_{uu}}$ are evaluated using the precomputed $k$-degree tensor $\mathcal{W}^{(3)}$, the squared polynomial is inherently truncated back to degree $k$. The higher-degree terms (degrees $k+1$ to $2k$) are omitted from the inner product, introducing a minor truncation error where the evaluated kurtosis slightly underestimates the exact full-tensor value.

For typical problems in astrodynamics, the magnitude of higher-order polynomial coefficients decays exponentially. Because the omitted terms contribute almost nothing to the overall statistical moment, the resulting truncation error is mathematically negligible (empirically observed to be on the order of $10^{-5}$ for the evaluated kurtosis). This minimal loss in accuracy is a highly favorable trade-off for the computational speedup and memory reduction achieved by bypassing the full fourth-order tensor contraction.

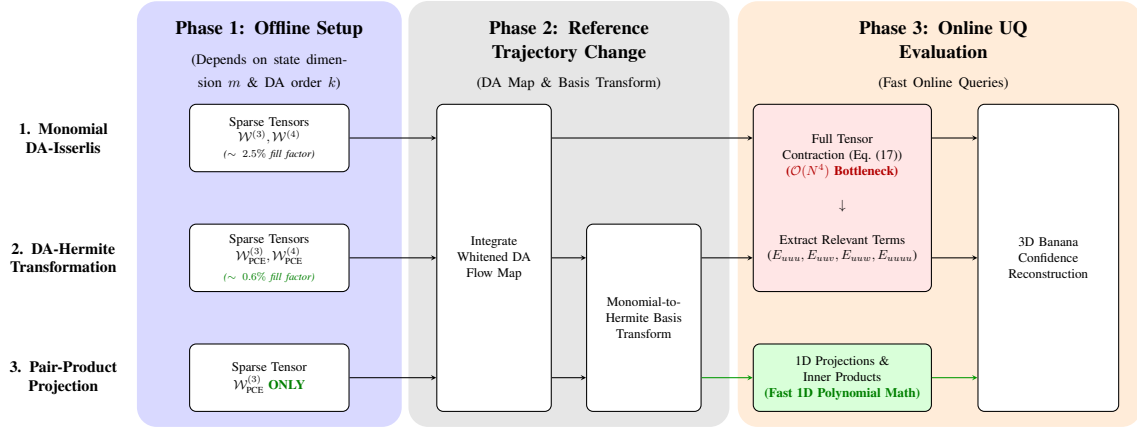
\begin{figure}[htbp]{
\centering
\resizebox{\textwidth}{!}{%
\begin{tikzpicture}[
    header/.style = {font=\Large\bfseries, text centered},
    methodHeader/.style = {font=\large\bfseries, text centered, text width=3.5cm},
    box/.style = {rectangle, rounded corners=5pt, draw=black, thick, fill=white, text centered, font=\small},
    arrow/.style = {thick, ->, >=stealth},
    background layer/.style={draw=none, rounded corners=15pt, fill opacity=0.15}
]

    \def\xLbl{-1.5}   
    \def\xOff{4.0}    
    \def\xDA{10.0}     
    \def\xHrm{14.0}   
    \def\xEval{19.3}  
    \def\xBan{24.8}   
    
    \def\yIss{0}      
    \def\yHer{-3.2}   
    \def\yPrj{-6.4}   

    \node[methodHeader] at (\xLbl, \yIss) {1. Monomial \\ DA-Isserlis};
    \node[methodHeader] at (\xLbl, \yHer) {2. DA-Hermite \\ Transformation};
    \node[methodHeader] at (\xLbl, \yPrj) {3. Pair-Product \\ Projection};

    \node[header, text width=6cm] at (\xOff, 2.2) (H1) {Phase 1: Offline Setup \\[0.15cm] \normalsize\mdseries (Depends on state dimension $m$ \& DA order $k$)};

    \node[box, text width=4cm, minimum height=1.8cm] (O1) at (\xOff, \yIss) {Sparse Tensors \\ $\mathcal{W}^{(3)}, \mathcal{W}^{(4)}$ \scriptsize \textit{($\sim 2.5\%$ fill factor)}};
    \node[box, text width=4cm, minimum height=1.8cm] (O2) at (\xOff, \yHer) {Sparse Tensors \\ $\mathcal{W}^{(3)}_{\text{PCE}}, \mathcal{W}^{(4)}_{\text{PCE}}$ \\[0.1cm] \scriptsize \textit{\textcolor{green!50!black}{($\sim 0.6\%$ fill factor)}}};
    \node[box, text width=4cm, minimum height=1.8cm] (O3) at (\xOff, \yPrj) {Sparse Tensor \\ $\mathcal{W}^{(3)}_{\text{PCE}}$ \textbf{\textcolor{green!50!black}{ONLY}}};

    \node[header, text width=7.5cm] at (12.0, 2.2) (H2) {Phase 2: Reference \\ Trajectory Change \\[0.15cm] \normalsize\mdseries (DA Map \& Basis Transform)};

    \node[box, text width=2.8cm, minimum height=8.2cm] (TDA) at (\xDA, \yHer) {Integrate \\ Whitened DA \\ Flow Map};

    \node[box, text width=2.8cm, minimum height=5.0cm] (THrm) at (\xHrm, -4.8) {Monomial-to-\\Hermite Basis \\ Transform \\};

    \node[header, text width=9.5cm] at (22.0, 2.2) (H3) {Phase 3: Online UQ \\ Evaluation \\[0.15cm] \normalsize\mdseries (Fast Online Queries)};

    \node[box, text width=4.5cm, minimum height=5.0cm, fill=red!10] (UCon) at (\xEval, -1.6) {
        Full Tensor \\ Contraction (Eq.\ \eqref{eq:tensor_contractions}) \\ 
        \textbf{\textcolor{red!70!black}{($\mathcal{O}(N^4)$ Bottleneck)}} \\[0.5cm] 
        $\downarrow$ \\[0.5cm] 
        Extract Relevant Terms ($E_{uuu}, E_{uuv}, E_{uuw}, E_{uuuu}$)
    };

    \node[box, text width=4.5cm, minimum height=1.8cm, fill=green!15] (UPrj) at (\xEval, \yPrj) {
        1D Projections \& \\ Inner Products \\ 
        \textbf{\textcolor{green!50!black}{(Fast 1D Polynomial Math)}}
    };

    \node[box, text width=3.5cm, minimum height=8.2cm] (UBan) at (\xBan, \yHer) {3D Banana \\ Confidence \\ Reconstruction};

    \draw[arrow] (O1.east) -- (TDA.west |- O1.east);
    \draw[arrow] (O2.east) -- (TDA.west |- O2.east);
    \draw[arrow] (O3.east) -- (TDA.west |- O3.east);

    \draw[arrow] (TDA.east |- O1.east) -- (UCon.west |- O1.east);
    
    \draw[arrow] (TDA.east |- O2.east) -- (THrm.west |- O2.east);
    \draw[arrow] (TDA.east |- O3.east) -- (THrm.west |- O3.east);

    \draw[arrow] (THrm.east |- O2.east) -- (UCon.west |- O2.east);
    \draw[arrow, thick, draw=green!60!black] (THrm.east |- O3.east) -- (UPrj.west |- O3.east);

    \draw[arrow] (UCon.east |- O1.east) -- (UBan.west |- O1.east);
    \draw[arrow] (UCon.east |- O2.east) -- (UBan.west |- O2.east);
    \draw[arrow, thick, draw=green!60!black] (UPrj.east |- O3.east) -- (UBan.west |- O3.east);

    \begin{scope}[on background layer]
        \node[fill=blue!15, rounded corners=15pt, fit=(H1) (O1) (O3), inner sep=0.4cm] {};
        \node[fill=black!10, rounded corners=15pt, fit=(H2) (TDA) (THrm), inner sep=0.4cm] {};
        \node[fill=orange!15, rounded corners=15pt, fit=(H3) (UCon) (UPrj) (UBan), inner sep=0.4cm] {};
    \end{scope}
\end{tikzpicture}
}}
\caption{Flowchart of the three main phases for the proposed analytical frameworks, highlighting the offline and online contributions.}
\label{fig:uq_flowchart}
\end{figure}

The flowchart in Figure \ref{fig:uq_flowchart} delineates the computational architecture of the three analytical approaches presented in Section \ref{sec:analytical_higher_order}. This three-phase structure isolates the heaviest computational burdens, offering specific advantages for onboard GNC and trajectory optimization applications.
First, the generation of the high-dimensional weight tensors (Phase 1) is strictly a function of the state dimension $m$ and the expansion order $k$. Because these tensors are entirely independent of the underlying dynamical system or the specific trajectory, they can be computed offline and stored in onboard memory. 
Second, the numerical integration of the dynamics via the DA flow map (Phase 2) is decoupled from the statistical parameters of the uncertainty. From a mission operations perspective, this means the map only needs to be recomputed when the spacecraft alters its nominal reference trajectory or when navigation errors exceed the polynomial radius of convergence. Consequently, engineers can rapidly evaluate thousands of different initial uncertainty distributions (e.g., during navigation filter tuning or covariance analysis) without ever re-integrating the equations of motion. Lastly, the online UQ evaluation (Phase 3) is restricted to lightweight polynomial algebra and 1D projections. This characteristic makes the "project-then-product" strategy highly suitable for robust trajectory optimization and stochastic guidance algorithms. Specifically, it provides the capability to evaluate accurate, non-Gaussian chance constraints, such as those required for strict collision avoidance or precision proximity operations, in real-time, without compromising onboard computational resources.

\section{Numerical Results}
\label{sec:numerical_results}
In this section, the proposed analytical framework is applied to representative scenarios in cislunar and small-body proximity regimes, where non-Gaussian uncertainty distributions naturally emerge. The performance of each method is evaluated in terms of computational cost and geometric accuracy against a Monte Carlo (MC) reference. A specific focus is placed on the framework's ability to efficiently capture non-elliptical confidence geometries via the analytical three-dimensional banana boundary reconstruction. 

The comparative analysis includes standard and DA-based implementations of LinCov, UT, GMM, and PCE, alongside the fully analytical higher-order moment extraction leveraging Isserlis' theorem, the Hermite transformation, and the pair-product projection strategy.

\subsection{Dynamical Environment}
\label{sec:dynamical_environment}
The dynamical environment is modeled to provide a medium-to-high-fidelity representation of the perturbations experienced in cislunar space and asteroid proximity operations. The perturbed $N$-body problem is formulated as \cite{michelotti_uncertainty_2024}:
\begin{equation}
\dot{\mathbf{r}}=\mathbf{v}, \qquad
\dot{\mathbf{v}}=\mathbf{a}_{\text{grav}}+\mathbf{a}_{\text{SRP}}+\mathbf{a}_{3B,\odot}+\mathbf{a}_{3B,\oplus}+\mathbf{a}_{3B,\text{Moon}}
\label{eq:dynamics}
\end{equation}

The first term, $\mathbf{a}_{\text{grav}} = \nabla U_{\text{grav}}$, encompasses the central-body gravitational acceleration. To account for irregular mass distributions, the gravitational potential is expanded via spherical harmonics \cite{vallado_book}:
\begin{equation}
    U_{\text{grav}} = \frac{\mu}{r} \sum_{l = 0}^{\infty} \sum_{m = 0}^{l} \left(\frac{R_e}{r}\right)^l \bar{P}_{l,m} ( \sin \phi)\left[ \bar{C}_{l,m} \cos m\lambda + \bar{S}_{l,m} \sin m\lambda \right]
    \label{eq:spherical_harmonics}
\end{equation}
where $\mu$ is the central body gravitational parameter, $R_e$ is the reference equatorial radius, and $\bar{P}_{l,m}$, $\bar{C}_{l,m}$, $\bar{S}_{l,m}$ are the associated Legendre polynomials and normalized harmonic coefficients, respectively. The resulting acceleration is subsequently rotated from the body-fixed frame to the inertial frame.

The $\mathbf{a}_{3B}$ terms represent the point-mass third-body perturbations from the Sun, Earth, and Moon. These are modeled as the differential Keplerian gravity acting between the central body and the spacecraft \cite{ferrariTrajectoryOptions}:
\begin{equation}
    \mathbf{a}_{3B,k} = \mu_k \left( \frac{\mathbf{r}_{k-\text{sc}}}{||\mathbf{r}_{k-\text{sc}}||^3} - \frac{\mathbf{r}_{k-\text{cb}}}{||\mathbf{r}_{k-\text{cb}}||^3} \right)
    \label{eq:third_body}
\end{equation}
where $\mu_k$ is the gravitational parameter of the $k$-th perturbing body, while $\mathbf{r}_{k-\text{sc}}$ and $\mathbf{r}_{k-\text{cb}}$ denote its relative position vector with respect to the spacecraft and the central body, respectively. 

Finally, $\mathbf{a}_{\text{SRP}}$ is the solar radiation pressure acceleration computed via a standard cannonball model \cite{scheeres2016orbital}:
\begin{equation}
    \mathbf{a}_{\text{SRP}} = -\frac{(1+\rho)P_0 A_{\text{sc}} AU}{c M_{\text{sc}}} \frac{\mathbf{r}_{\text{sc-Sun}}}{||\mathbf{r}_{\text{sc-Sun}}||^3}
    \label{eq:srp_cannonball}
\end{equation}
where $P_0$ is the solar flux at 1 AU, $c$ is the speed of light, and the spacecraft is characterized by its mass $M_{\text{sc}}$, cross-sectional area $A_{\text{sc}}$, and reflectance coefficient $\rho$. The values considered in this work are related to a representative spacecraft which resembles a 6U Deep Space CubeSat inspired by Milani CubeSat ($M_{sc} = 12$ kg, $A_{sc} = 0.5 $m$^2$, $\rho = 0.3$) \footnote{\url{https://www.heramission.space/hera-mission-milani-cubesat}. Last accessed July 1, 2026.}.

The planetary and asteroidal ephemerides required to evaluate these perturbations are retrieved using NAIF SPICE kernels and JPL Horizons system.

\subsection{Cislunar NRHO}
\label{sec:NRHO}
Near-Rectilinear Halo Orbits (NRHOs) around the Earth-Moon $L_2$ point are of critical interest for current and future cislunar architectures. In this scenario, uncertainty propagation is analyzed along a 1.5 days orbital arc passing through the perilune. The initial state, extracted from the Lunar Gateway SPICE kernel\footnote{\url{https://naif.jpl.nasa.gov/pub/naif/misc/MORE_PROJECTS/DSG/}. Last accessed July 1, 2026.}, and the initial Gaussian distribution are detailed in Table \ref{tab:initial_conditions_NRHO}. The initial uncertainty is deliberately scaled to exceed nominal cislunar navigation baselines, of comparable order to stress-case dispersions used to force nonlinear covariance growth in CR3BP-based cislunar trajectory studies \cite{kelly2024robust}. This serves as a computational stress test, designed to force the rapid manifestation of severe nonlinear spatial deformations over a short propagation window.
The Moon's gravitational spherical harmonics up to fourth-order are included for this analysis.

\begin{table}[htbp]
\fontsize{10}{10}\selectfont
\caption{Initial state and uncertainty for Cislunar NRHO scenario.}
\label{tab:initial_conditions_NRHO}
\centering
\begin{tabular}{l|c}
\hline
Scenario & Cislunar NRHO \\
\hline
Init. Position $\mathbf{r}_0$ (km) & $\begin{bmatrix} -4.731\times10^{2} & 1.897\times10^{4} & -3.546\times10^{4} \end{bmatrix}^{T}$ \\
Init. Velocity $\mathbf{v}_0$ (km/s) & $\begin{bmatrix} 1.26\times10^{-2} & -3.64\times10^{-2} & 3.60\times10^{-1} \end{bmatrix}^{T}$ \\
Pos. Std $\boldsymbol{\sigma}_{\mathbf{r}_0}$ (km) & $diag\begin{pmatrix} 10 & 1 & 1 \end{pmatrix}$ \\
Vel. Std $\boldsymbol{\sigma}_{\mathbf{v}_0}$ (km/s) & $diag\begin{pmatrix} 15\times10^{-3} & 1\times10^{-3} & 1\times10^{-3} \end{pmatrix}$ \\
\hline
\end{tabular}
\end{table}

\begin{figure}[htbp]
    \centering
    \subfigure{\includegraphics[width=0.45\textwidth]{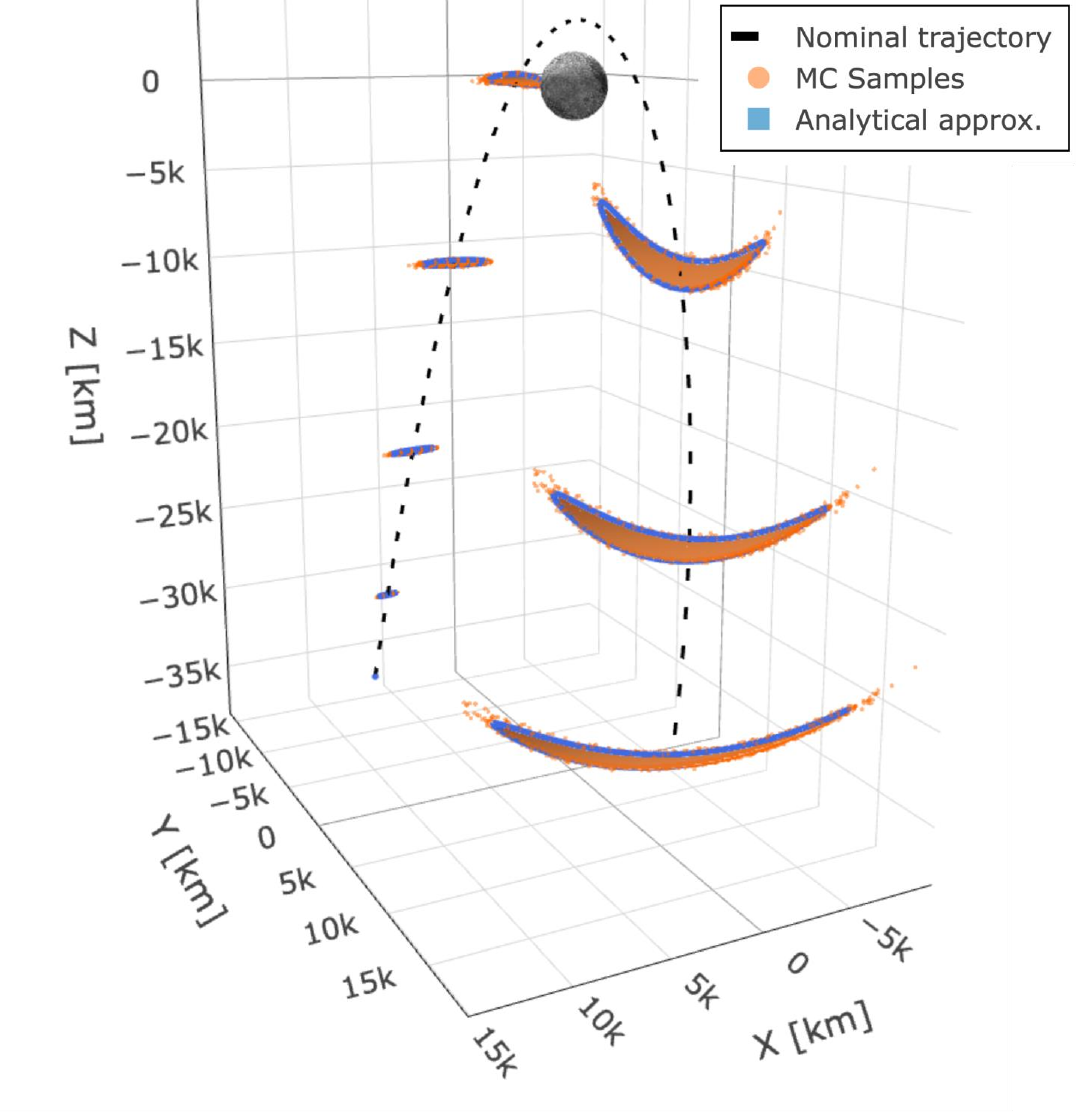}}
    \hfill
    \subfigure{\includegraphics[width=0.45\textwidth]{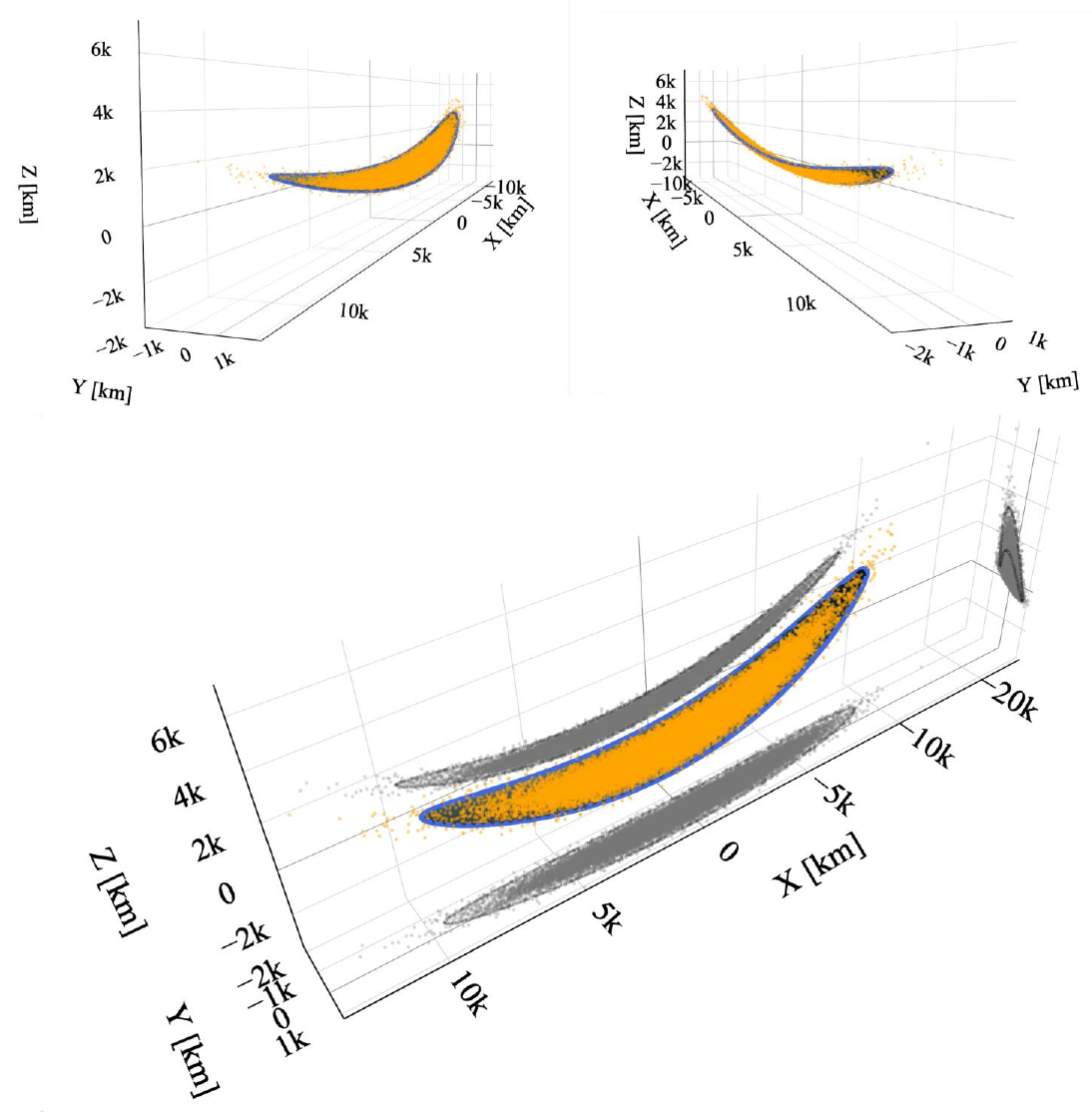}}
    \caption{Cislunar NRHO scenario. Left: Nominal trajectory with uncertainty propagation distribution at different instants. Right: Different viewing angles of the analytical boundary reconstruction informed by higher moments at final time.}
    \label{fig:NRHO_traj}
\end{figure}

Figure \ref{fig:NRHO_traj} illustrates the nominal trajectory and the evolution of the uncertainty distribution. As the spacecraft approaches perilune, the strong local gravitational gradient induces shear across the probability density function. The initial small Gaussian distribution is rapidly stretched into a non-Gaussian one, showing the typical banana shape.
Note that the cross-track orientation of the banana is induced by the asymmetrical initial covariance, which is characterized by larger uncertainties along the x direction.
The right panel of Figure \ref{fig:NRHO_traj} focuses on the final time instant,  overlaying the analytical 3D banana surface reconstruction against the MC samples. Different viewing angles are shown to highlight the capability of the analytical approximation to accurately capture the bending of the distribution in the entire three-dimensional space, both in-plane and out-of-plane.

\begin{figure}[htbp]
    \centering
    \subfigure{\includegraphics[width=0.75\textwidth]{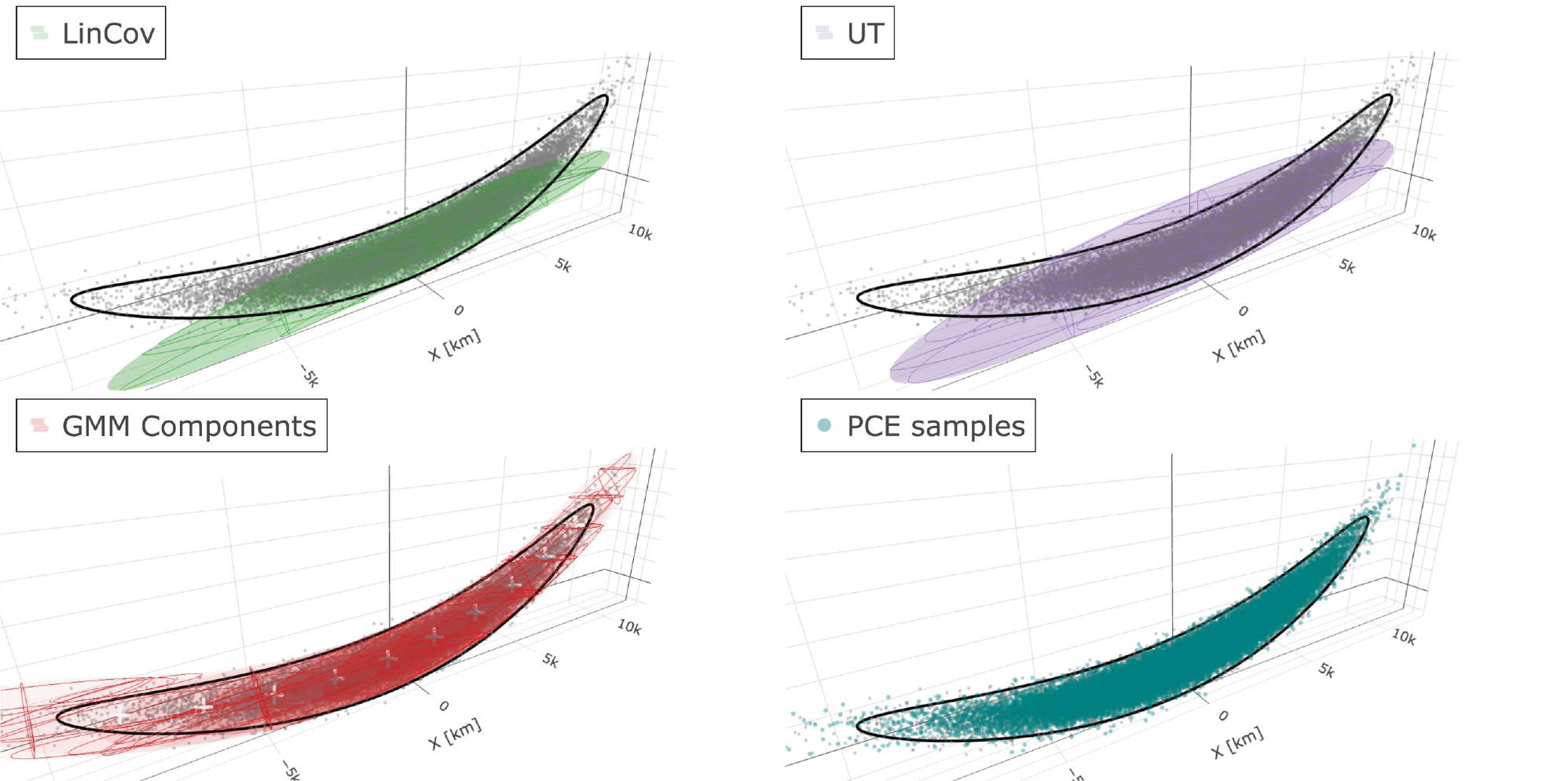}}
    \caption{Comparison of different UQ methods at the final time instant, with respect to the MC solution (gray dots) and the DA analytical approximation (black solid line).}
    \label{fig:uq_comparison_NRHO}
\end{figure}

Figure \ref{fig:uq_comparison_NRHO} compares the final propagated distributions obtained using standard astrodynamics UQ methods: LinCov, UT, GMM, and PCE. The traditional covariance ellipsoids produced by LinCov and UT do not capture the extended tails of the actual final distribution. The GMM is implemented using 10 Gaussian components distributed along the principal axis of the initial uncertainty with a bell-shaped weight distribution. In the plot, this model is visualized by the individual component ellipsoids, with opacity proportional to their respective weights. This fixed-axis splitting strategy yields a low-fidelity approximation of the highly non-Gaussian shape, in particular regarding the approximation of higher-order moments; achieving higher accuracy would require a more computationally expensive recursive splitting algorithm. In contrast, the 4th-order non-intrusive PCE accurately reconstructs the full non-Gaussian shape of the distribution with high accuracy.

Table \ref{tab:timings_accuracy_NRHO} summarizes the computational runtimes and statistical moment accuracies for the evaluated methods. All simulations were performed on a 2021 MacBook Pro (Apple M1 chip, 16 GB RAM). Relative errors are computed against a reference Monte Carlo (MC) simulation of $N=25{,}000$ samples, which required $503.15$ s for numerical integration. Expanding the DA flow map to the 4th order required an offline computational time of $10.27$ s. 
Note that the computational cost of the DA integration is highly dependent on the degree and order of the spherical harmonics gravity model, as higher degrees significantly increase the number of polynomial operations required per integration step. Consequently, the computational overhead of DA integration relative to standard floating-point integration widens as the gravity model fidelity increases.

Covariance errors are computed using the Frobenius norm of the covariance matrix. Skewness and kurtosis errors are instead computed as the relative norm of the marginal moments,
\begin{equation}
    \varepsilon_{\gamma} = \frac{\left\lVert \boldsymbol{\gamma}_{\text{method}} - \boldsymbol{\gamma}_{\text{MC}} \right\rVert}{\left\lVert \boldsymbol{\gamma}_{\text{MC}} \right\rVert}, \qquad
    \varepsilon_{\kappa} = \frac{\left\lVert \boldsymbol{\kappa}_{\text{method}} - \boldsymbol{\kappa}_{\text{MC}} \right\rVert}{\left\lVert \boldsymbol{\kappa}_{\text{MC}} \right\rVert},
    \label{eq:skew_kurt_error}
\end{equation}
where $\boldsymbol{\gamma}$ and $\boldsymbol{\kappa}$ denote the vectors of marginal skewness and kurtosis, extracted from the diagonal of the respective third- and fourth-order moment tensors, and the subscript MC denotes the Monte Carlo reference.
For the DA ``project-then-product'' approach, the reported skewness and kurtosis errors instead correspond to the mean of the errors of each of the extracted components (e.g., $E_{uuu}$, $E_{uuv}$, $E_{uuw}$, $E_{uuuu}$).

\begin{table}[htbp]
    \centering
    \caption{Comparison of computational runtime and relative accuracy against standard Monte Carlo ($25,000$ samples) at final time $t_f$.}
    \label{tab:timings_accuracy_NRHO}
    
    \renewcommand{\arraystretch}{1.2} 
    
    \resizebox{\textwidth}{!}{%
    \begin{tabular}{l cc cccc}
        \toprule
        & \multicolumn{2}{c}{\textbf{Runtime [s]}} & \multicolumn{4}{c}{\textbf{Relative Error vs. Monte Carlo}} \\
        \cmidrule(lr){2-3} \cmidrule(l){4-7}
        \textbf{Method} & \textbf{Online Eval.} & \textbf{Total} & \textbf{Mean} & \textbf{Covariance} & \textbf{Skewness} & \textbf{Kurtosis} \\
        \midrule
        
        \multicolumn{7}{l}{\textit{Standard Frameworks}} \\
         LinCov                               &  0.2496 &  0.2496 & $7.34 \times 10^{-3}$  & $1.36 \times 10^{-2}$  & --                     & -- \\
        UT                                   &  0.2726 &  0.2726 & $6.26 \times 10^{-4}$  & $3.30 \times 10^{-3}$  & --                     & -- \\
        PCE (p=4)                         &  9.3443 &  9.4864 & $6.16 \times 10^{-4}$  & $1.41 \times 10^{-2}$  & $2.20 \times 10^{-2}$  & $3.49 \times 10^{-2}$ \\
        GMM (K=10)                        &  2.6434 &  2.6434 & $6.17 \times 10^{-4}$  & $1.13 \times 10^{-2}$  & $1.51 \times 10^{-1}$  & $1.34 \times 10^{-1}$ \\

       Monte Carlo                          & 503.1478 & 503.1478 & Ref.          & Ref.          & Ref.          & Ref. \\
        \midrule
        \multicolumn{7}{l}{\textit{Differential Algebra Frameworks}} \\
        \textbf{DA Full Tensors (4th order)} &  5.5289 & 17.1013 & $6.17 \times 10^{-4}$  & $1.22 \times 10^{-2}$  & $1.83 \times 10^{-2}$  & $5.49 \times 10^{-2}$ \\
        \textbf{DA Project-then-product (4th order)} &  0.4868 & 12.0592 & $6.17 \times 10^{-4}$  & $1.22 \times 10^{-2}$  & $2.31 \times 10^{-2}$  & $2.53 \times 10^{-2}$ \\

        \bottomrule
    \end{tabular}%
    }
    
    \vspace{0.15cm}
    \raggedright
    \footnotesize{\textit{Note:} Total runtime for DA frameworks includes the expansion of the whitened DA flow map. Skewness and kurtosis moments are inherently not extracted for standard LinCov and UT frameworks.}
\end{table}

\begin{figure}[htbp]
    \centering
    \subfigure{\includegraphics[width=\textwidth]{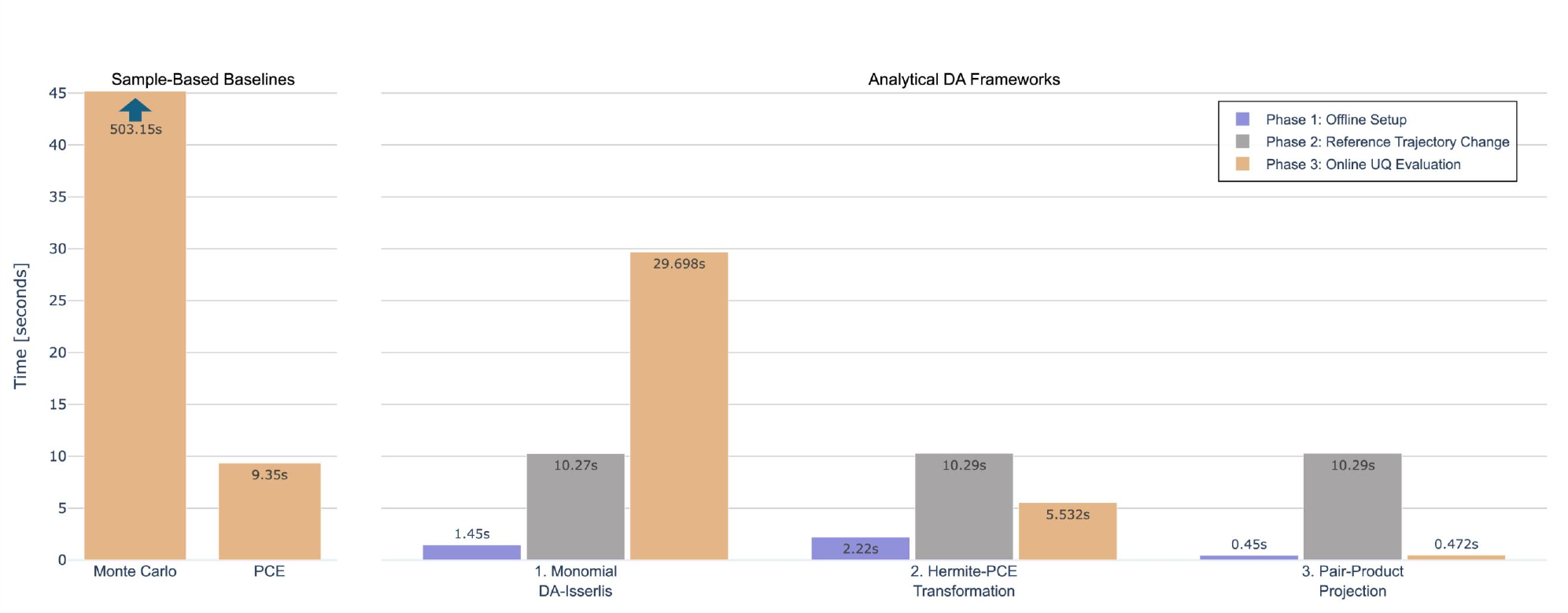}}
    \caption{Runtime breakdown for analytical higher-order moments recovery approaches using DA framework.}
    \label{fig:time_breakdown_NRHO}
\end{figure}

The fully analytical higher-order moments are extracted by exploiting the 4th-order DA flow map. The computational performance of the different approaches described in Section \ref{sec:analytical_higher_order} is detailed in Figure \ref{fig:time_breakdown_NRHO}, which illustrates the runtime breakdown for higher-order moment recovery without sampling. 
Because the Isserlis and Hermite weight tensors are entirely independent of the orbital dynamics, they can be computed offline and stored. Consequently, the onboard computational requirement is strictly limited to the DA map expansion, when necessary for updating the reference trajectory, and the subsequent tensor contractions for the uncertainty evaluation. By leveraging the inherent sparsity of the Hermite basis transformation (e.g., a 0.6\% fill factor), the online computational burden is significantly reduced compared to applying Isserlis' theorem directly to the standard monomial basis.
Finally, the project-then-product strategy enables the targeted computation of only the specific skewness and kurtosis terms required to parameterize the confidence boundary, entirely bypassing the need for full tensor contractions or discrete sample evaluations. 

Note that the reduction in offline computation time achieved by the project-then-product approach stems directly from the pair-product strategy detailed in Section \ref{sec:pair-product}. By enabling the derivation of fourth-order kurtosis terms through the contraction of lower-order tensors, this strategy eliminates the necessity of computing the more expensive four-dimensional kurtosis tensor.
Conversely, the marginal increase in expansion time for the Hermite and pair-product methods is due to the necessary basis conversion from the monomial representation to the Hermite basis.

The computational advantage of the project-then-product method over full-tensor approaches becomes increasingly evident as the DA expansion order increases. In DA, the computational effort scales combinatorially: for instance, in a system with $n=6$ dimensions, a second-order expansion requires the integration of 258 equations, whereas a third-order expansion necessitates 1554. This scaling is intensified when extracting higher-order statistical moments; because moments are computed via the expectation of products of state variables, the underlying monomial degree grows rapidly. Specifically, resolving $m$-th order moments from a $k$-th order DA map requires tracking monomials up to degree $m \times k$. For example, resolving 4th-order kurtosis components from a 3rd-order DA map requires order 12, whereas from a 4th-order map it necessitates order 16. By projecting the dynamics onto the principal axes, the project-then-product method bypasses the need to construct the entire high-degree tensors, mitigating the dimensionality bottleneck.

\begin{figure}[htbp]
    \centering
    \subfigure{\includegraphics[width=\textwidth]{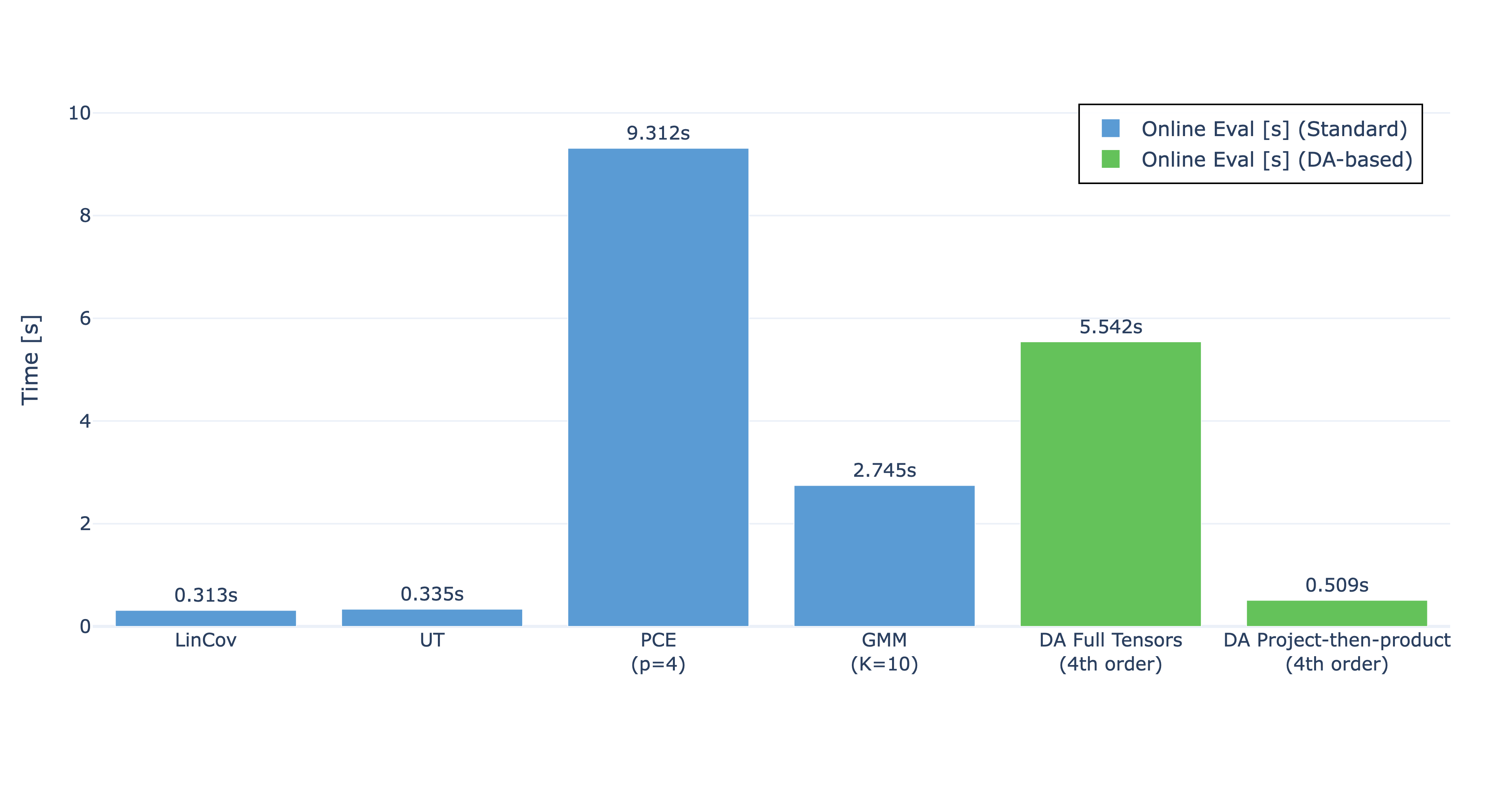}}
    \caption{Computational times of different UQ methods for cislunar scenario. Note that DA methods times do not include the DA expansion time, but are related to the single UQ query. DA Full Tensors refers to the faster Hermite-DA approach.}
    \label{fig:time_NRHO}
\end{figure}

\begin{figure}[htbp]
    \centering
    \subfigure{\includegraphics[width=\textwidth]{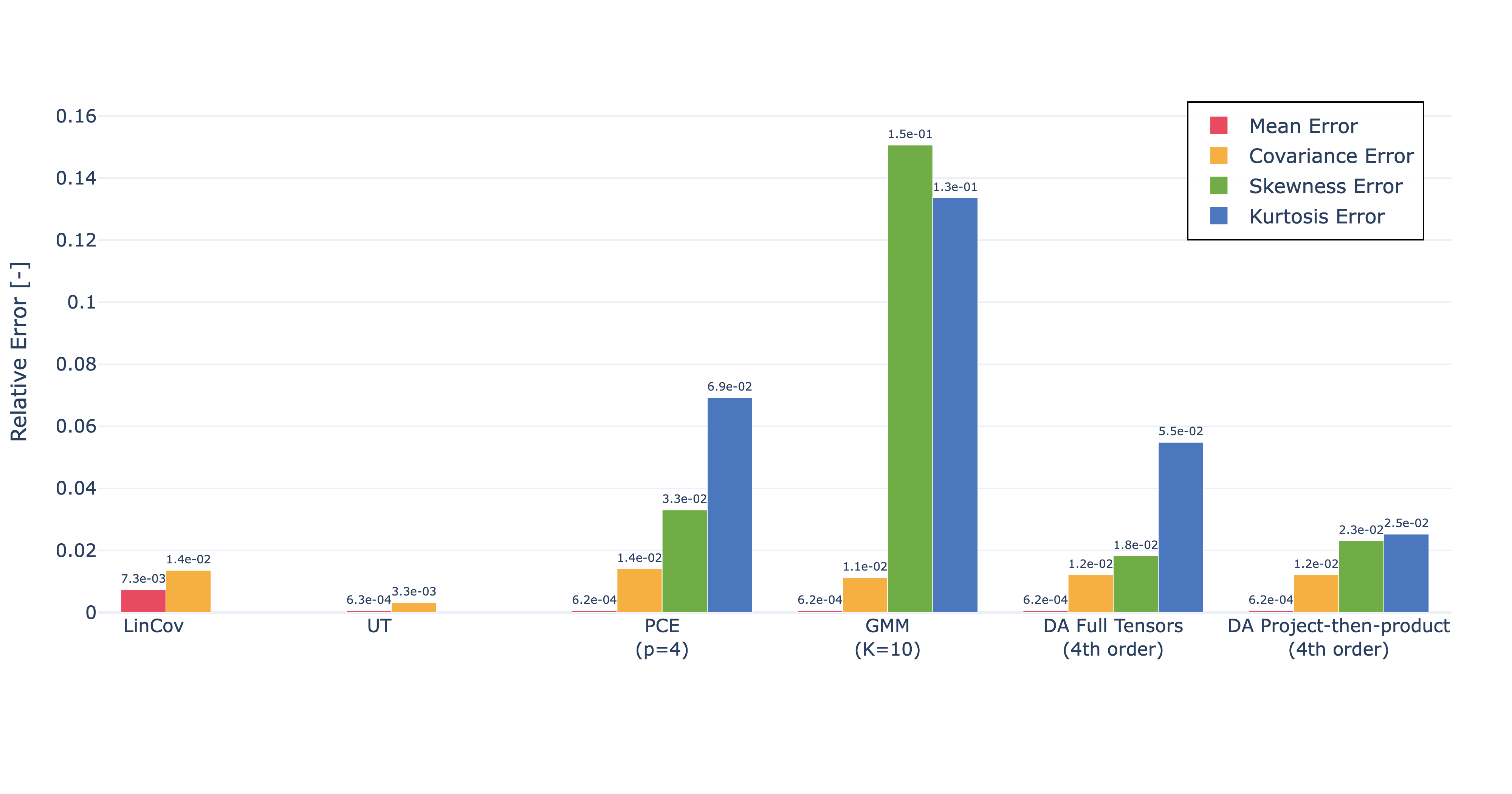}}
    \caption{Statistical moments relative errors with respect to the MC reference for different UQ methods for cislunar scenario. Note that the error of skewness and kurtosis of DA project-then-product approach refers only to the extracted terms.}
    \label{fig:err_NRHO}
\end{figure}

Results of the performance comparison with state-of-the-art methods are also shown in Figures \ref{fig:time_NRHO} and \ref{fig:err_NRHO}. Figure \ref{fig:time_NRHO} details the computational time for a single UQ query. For the DA-analytical approaches, this online evaluation time excludes the offline DA map expansion, as a new expansion is only required when the estimated truncation error of the polynomial map exceeds a prescribed threshold. Figure \ref{fig:err_NRHO} presents the relative errors of the statistical moments compared to the Monte Carlo (MC) ground truth. Note that the GMM method implemented in this work does not actively optimize the splitting of the Gaussian components and their weights, and therefore it shows low computational time but relatively high error with respect to other high-fidelity methods. Together, these plots demonstrate that the DA-analytical approaches estimate the mean, covariance, and higher-order moments with an accuracy comparable to high-fidelity methods like PCE.  Furthermore, the project-then-product strategy significantly speeds up the online UQ queries, achieving runtimes similar to lower-fidelity methods such as LinCov and UT. Therefore, this reduction in computational overhead is achieved while maintaining the geometric accuracy of non-Gaussian high-fidelity methods. The analytically recovered marginal skewness and kurtosis closely match the fully integrated MC reference, validating the formulation of the Hermite basis transformation and the pair-product strategy in highly nonlinear regimes.

\subsection{Apophis Proximity Trajectories}
\label{sec:apophis}
To further validate the proposed framework, similar analyses are conducted for close-proximity operations around the asteroid Apophis. This environment presents unique UQ challenges driven by the asteroid's highly irregular gravity field, relevant Solar Radiation Pressure, and rapid dynamical timescales. Two specific scenarios are evaluated: a revolution during deep-space proximity operations, and a highly perturbed trajectory during the asteroid's predicted 2029 close approach to Earth\cite{michelotti_uncertainty_2024}. Apophis’s gravity spherical harmonics to the 4th order
are used for this analysis.

The initial state vectors are detailed in Table \ref{tab:initial_conditions_apophis}. For both scenarios, the initial Gaussian distribution is parameterized by a spherical position variance of $\sigma_r = 10$ m and spherical velocity variance of $\sigma_v = 0.3 \times 10^{-3}$ m/s. The propagation time horizon is set to 2 days for the deep-space scenario and 0.75 days for the accelerated Earth flyby scenario. As in the previous case study, the reference Monte Carlo simulations utilizes $N=25{,}000$ samples, and the DA map is expanded to the 4th-order.

\begin{table}[htbp]
\fontsize{10}{10}\selectfont
\caption{Initial state for Apophis proximity scenarios.}
\label{tab:initial_conditions_apophis}
\centering
\begin{tabular}{l|c}
\hline
Scenario & Apophis deep-space \\
\hline
Init. Position $\mathbf{r}_0$ (km) & $\begin{bmatrix} 0.85 & 0 & 0 \end{bmatrix}^{T}$ \\
Init. Velocity $\mathbf{v}_0$ (km/s) & $\begin{bmatrix} 2.7229\times10^{-5} & -2.1036\times10^{-5} & 4.1277\times10^{-5} \end{bmatrix}^{T}$ \\
\hline
Scenario & Apophis Earth's flyby \\
\hline
Init. Position $\mathbf{r}_0$ (km) & $\begin{bmatrix} -0.3255 & -1.4633 & -0.0520 \end{bmatrix}^{T}$ \\
Init. Velocity $\mathbf{v}_0$ (km/s) & $\begin{bmatrix} -2.8502\times10^{-5} & 1.9168\times10^{-5} & -0.1889\times10^{-5} \end{bmatrix}^{T}$ \\
\hline
\end{tabular}
\end{table}

\begin{figure}[htbp]
    \centering
    \subfigure{\includegraphics[width=0.48\textwidth]{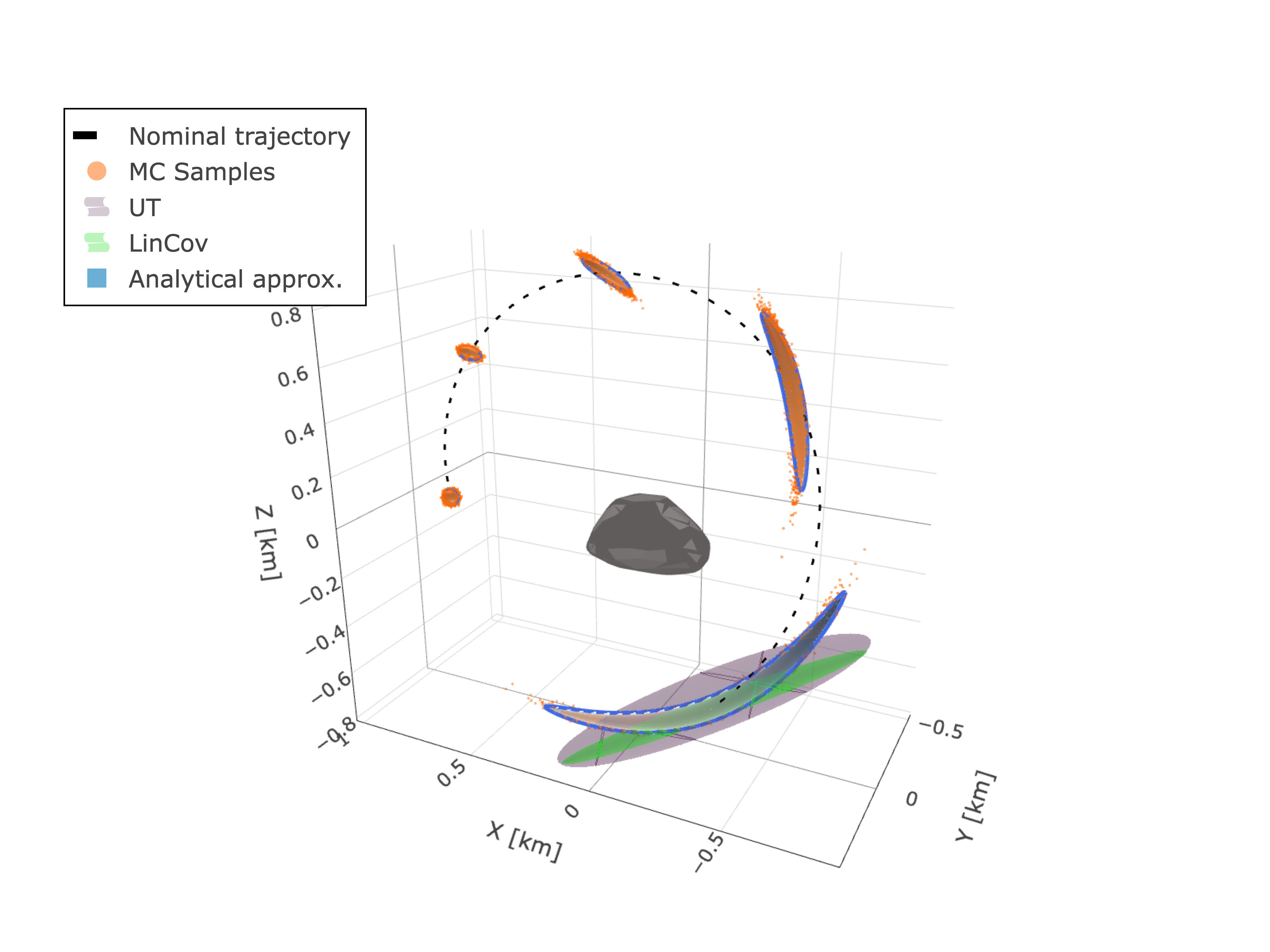}}
    \hfill
    \subfigure{\includegraphics[width=0.48\textwidth]{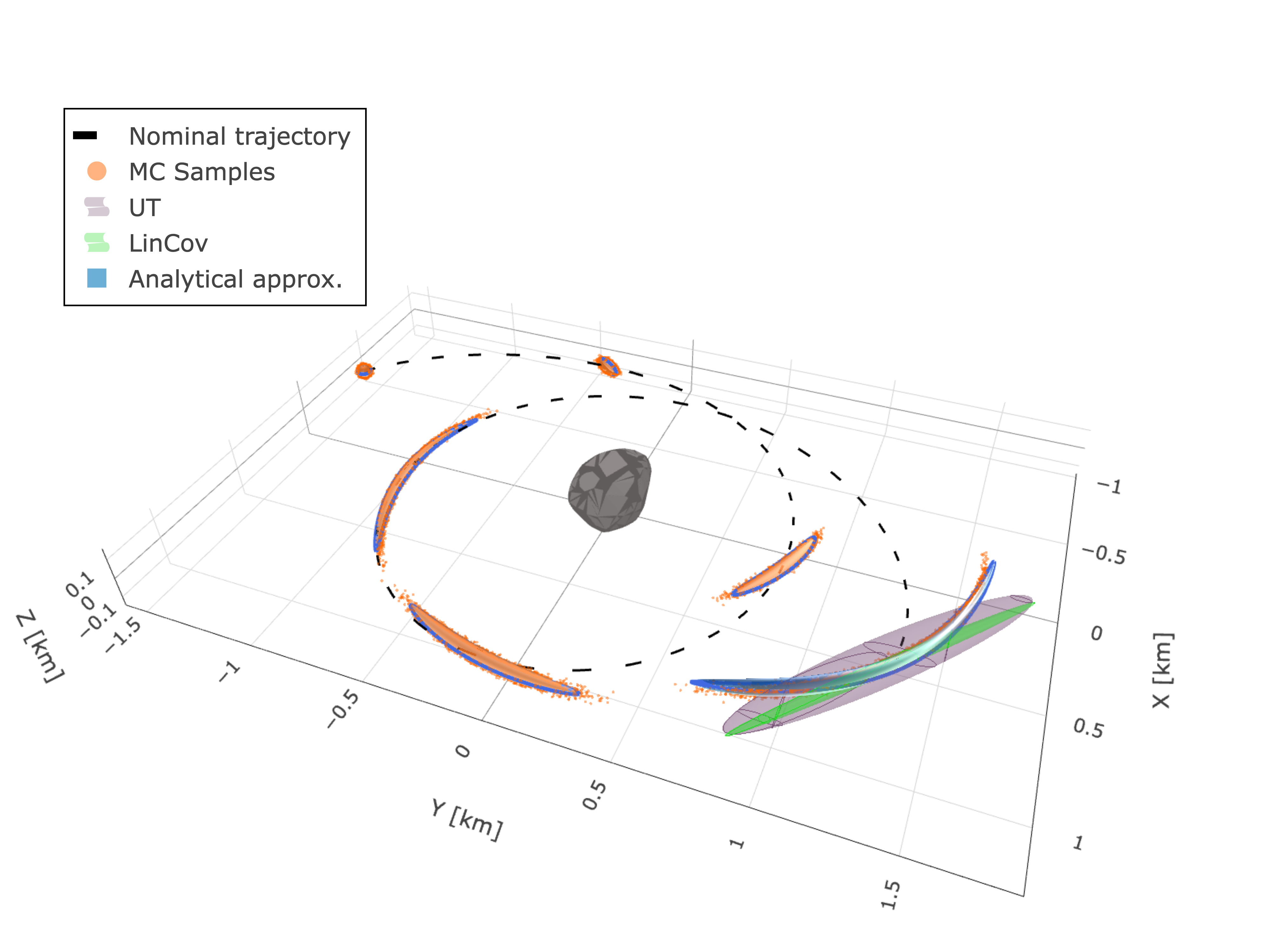}}
    \caption{Apophis proximity scenarios. Left: During deep-space. Right: During Earth's flyby.}
    \label{fig:apophis_traj}
\end{figure}

The resulting uncertainty propagation is illustrated in Figure \ref{fig:apophis_traj}. In both cases, the complex and perturbed dynamics cause the initial Gaussian sphere to shear into a non-elliptical volume. The plots overlay the nominal trajectory, the sample distributions, and the corresponding 3D banana boundary reconstruction at the final time instant. The analytical approximation, parameterized by the analytically recovered skewness and kurtosis, successfully envelopes the non-Gaussian features of the distribution, providing a strict geometric bound that standard covariance ellipsoids fail to capture.
In three-dimensional space, the theoretical containment fraction for a $3\sigma$ Gaussian confidence boundary is $97.07\%$. The LinCov ellipsoid includes only $69.43\%$ of the MC samples, significantly underestimating the true dispersion. The UT contains $94.54\%$ of the samples, achieving a closer containment fraction as expected from a nonlinear method. However, because UT still restricts the confidence boundary to a Gaussian distribution, it completely misses its spatial curvature and includes large regions of empty space while excluding the extended tails. In contrast, the proposed analytical approximation accurately reconstructs the non-Gaussian banana shape, containing $96.28\%$ of the samples.

\begin{figure}[htbp]
    \centering
    \subfigure{\includegraphics[width=\textwidth]{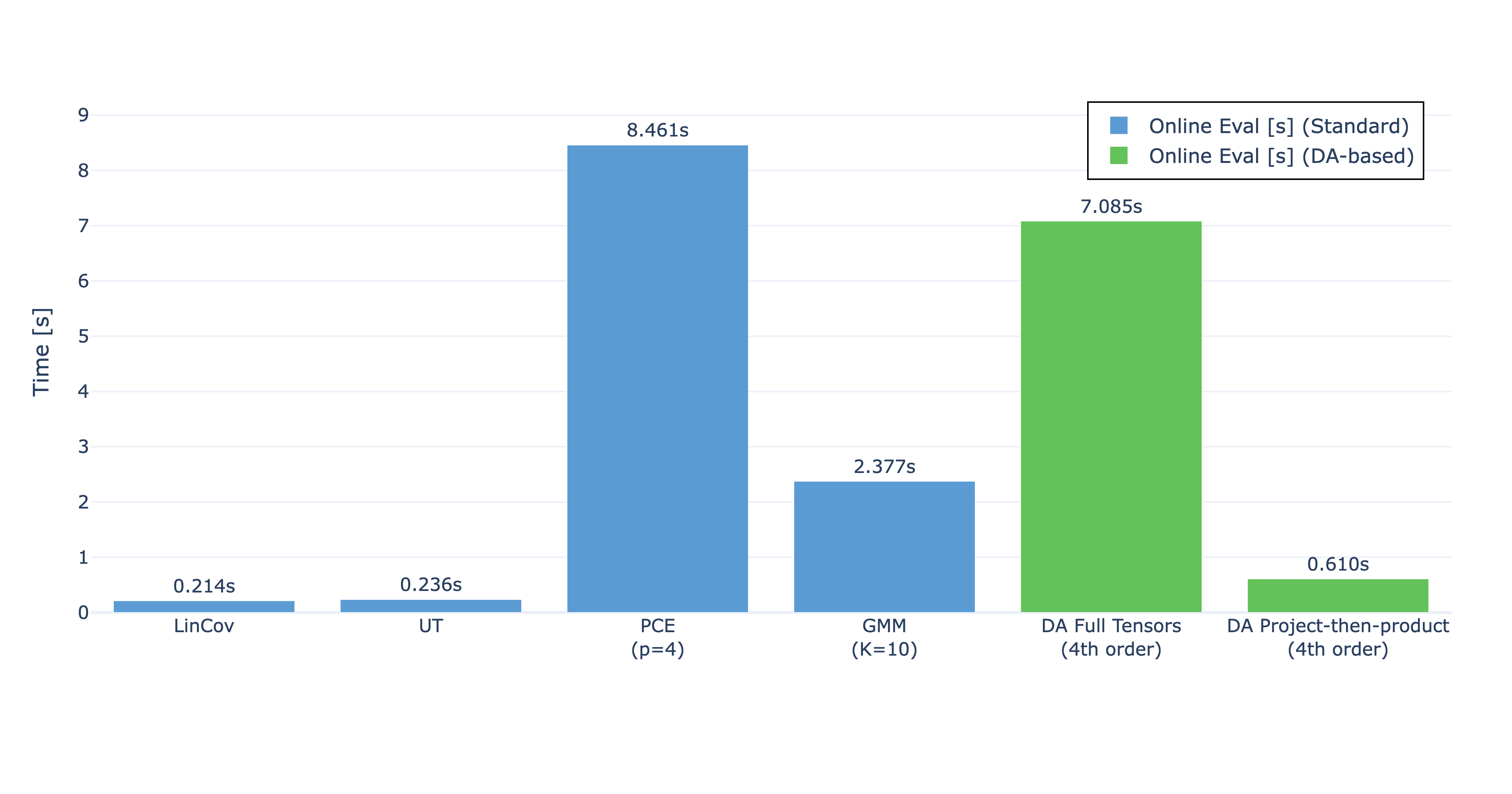}}
    \caption{Computational times of different UQ methods for Apophis (Earth's flyby scenario). Note that DA methods times do not include the DA expansion time, but are related to the single UQ query. DA Full Tensors refers to the faster Hermite-DA approach.}
    \label{fig:time_apophis}
\end{figure}

\begin{figure}[htbp]
    \centering
    \subfigure{\includegraphics[width=\textwidth]{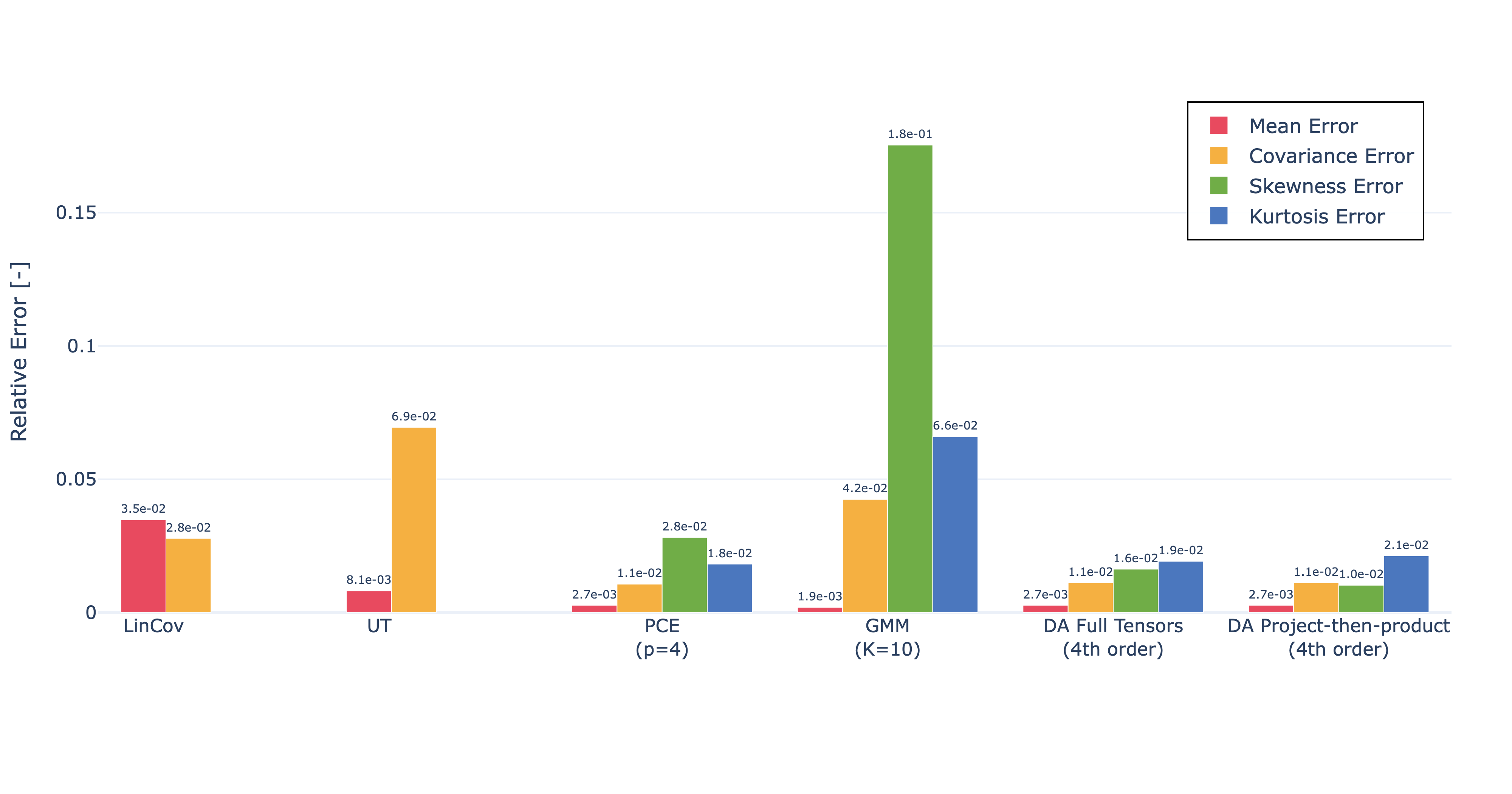}}
    \caption{Statistical moments relative errors with respect to the MC reference for different UQ methods for Apophis (Earth's flyby scenario). Note that the error of skewness and kurtosis of DA project-then-product approach refers only to the extracted terms.}
    \label{fig:err_apophis}
\end{figure}

The computational times of the standard UQ methods and DA-based analytical implementations for the Earth's flyby scenario are summarized in Figure \ref{fig:time_apophis}. 
The relative errors, with respect to the reference MC simulation, of the propagated statistical moments at the final time instants are presented in Figure \ref{fig:err_apophis}. The results confirm that the DA-based analytical implementations maintain geometric and statistical accuracy equivalent to their standard numerical counterparts, validating that the profound reduction in computational time does not compromise the fidelity of the uncertainty quantification.

\section{Conclusion}
\label{sec:conclusion}
This work introduces an analytical framework for non-Gaussian uncertainty propagation and three-dimensional confidence boundary reconstruction in highly perturbed dynamical environments. To mitigate the high computational cost and statistical noise of traditional sample-based methods, the proposed methodology leverages a fully analytical approach to extract statistical moments. A differential algebra (DA) formulation is used to replace the numerical integration of the dynamics with high-order polynomial flow maps. Then, higher-order moments are extracted from the DA coefficients (i.e., without sampling) by exploiting Isserlis' theorem and the mapping of standard monomials into an orthogonal probabilist's Hermite basis, which significantly improves the tensor sparsity. 
Additionally, a ``project-then-product'' approach is used to bypass the computational and memory scaling associated with high-order moments tensor contractions, and to retrieve only the relevant terms. The analytically extracted skewness and kurtosis are then utilized to parameterize a three-dimensional analytical banana-shaped confidence boundary. This geometric reconstruction accurately bounds the spatial, bending, and out-of-plane torsion caused by propagation in nonlinear dynamics, which traditional linear methods fail to capture. It also has the benefit of being analytic, explicitly parameterized by angles. 

The proposed framework is tested and validated in two challenging nonlinear regimes: a dynamically realistic cislunar near-rectilinear halo orbit (NRHO), and close-proximity trajectories around the asteroid Apophis in deep space and during Earth flyby. The numerical results demonstrate that the fully analytical approach, exploiting tensor sparsity and projection techniques, achieves geometric and statistical accuracy equivalent to a large-batch Monte Carlo study, or other high-fidelity uncertainty propagation methods, while simultaneously reducing computational times. This sample-free methodology provides an efficient and mathematically rigorous tool for onboard uncertainty quantification (UQ), robust trajectory optimization, and space domain awareness for modern astrodynamics applications. The methodology can be augmented with trajectory optimization techniques making use of polynomial maps like DA or the related state transition tensors (STTs),\cite{BurnTopp2025,RegantiniJGCD,caleb2026daddy} whereby the same maps that are used for nominal trajectory correction can be used for nonlinear UQ operations.

\bibliographystyle{AAS_publication}
\bibliography{references.bib}

@article{RegantiniJGCD,
	author = {Regantini, Omar and Burnett, Ethan R. and Rizza, Antonio and Morselli, Alessandro and Topputo, Francesco},
	doi = {10.2514/1.G009332},
	journal = {Journal of Guidance, Control, and Dynamics},
	pages = {1-15},
	title = {Convex Trajectory Optimization via Monomial Coordinates Transcription for Cislunar Rendezvous},
	year = {2026}}

@article{caleb2026daddy,
      title={Taylor polynomial-based constrained solver for fuel-optimal low-thrust trajectory optimisation}, 
      author={Thomas Caleb and Roberto Armellin and Spencer Boone and Stéphanie Lizy-Destrez},
      year={2026},
      eprint={2502.00398},
      archivePrefix={arXiv},
      primaryClass={math.OC},
      doi={10.48550/arXiv.2502.00398}, 
}

@article{BurnTopp2025,
	author = {Burnett, Ethan R. and Topputo, Francesco},
	doi = {10.2514/1.G008512},
	journal = {Journal of Guidance, Control, and Dynamics},
	number = {4},
	pages = {736-756},
	title = {Rapid Nonlinear Convex Guidance Using a Monomial Method},
	volume = {48},
	year = {2025}}

@conference{AAS26a,
	author = {Burnett, Ethan R. and Spencer Boone and Niccol{\`o} Michelotti},
	booktitle = {AAS/AIAA Astrodynamics Specialist Conference},
	month = {July},
	number = {AAS 26-899},
	organization = {American Astronautical Society},
	title = {An Efficient Non-Gaussian Chance Constraint Method for Stochastic Nonlinear Problems in Spaceflight},
	year = {2026}}

@article{acciarini_nonlinear_2025,
	title = {Nonlinear {Propagation} of {Non}-{Gaussian} {Uncertainties}},
	volume = {48},
	issn = {0731-5090, 1533-3884},
	url = {https://arc.aiaa.org/doi/10.2514/1.G008717},
	doi = {10.2514/1.G008717},
	language = {en},
	number = {4},
	urldate = {2026-02-06},
	journal = {Journal of Guidance, Control, and Dynamics},
	author = {Acciarini, Giacomo and Baresi, Nicola and Lloyd, David J. B. and Izzo, Dario},
	month = apr,
	year = {2025},
	pages = {903--913},
}

@article{valli_nonlinear_2013,
	title = {Nonlinear {Mapping} of {Uncertainties} in {Celestial} {Mechanics}},
	volume = {36},
	issn = {0731-5090, 1533-3884},
	url = {https://arc.aiaa.org/doi/10.2514/1.58068},
	doi = {10.2514/1.58068},
	language = {en},
	number = {1},
	urldate = {2026-02-27},
	journal = {Journal of Guidance, Control, and Dynamics},
	author = {Valli, M. and Armellin, R. and Di Lizia, P. and Lavagna, M. R.},
	month = jan,
	year = {2013},
	pages = {48--63},
}

@inproceedings{BurnettBooneCDC26,
	author = {E. Burnett and S. Boone},
	booktitle = {IEEE Conference on Decision and Control},
	title = {Analytic Non-Gaussian Chance Constraint Method for Stochastic Trajectory Control},
	year = {Preprint pending appearance on arXiv in 4/2026.}}

@article{azor1982combinatorial,
  title={Combinatorial applications of Hermite polynomials},
  author={Azor, Ruth and Gillis, Joseph and Victor, Jonathan D},
  journal={SIAM Journal on Mathematical Analysis},
  volume={13},
  number={5},
  pages={879--890},
  year={1982},
  publisher={SIAM}
}

@article{michelotti_uncertainty_2024,
	title = {Uncertainty {Propagation} {Performance} in {Proximity} {Operations} {Around} {Small} {Bodies}},
	volume = {71},
	issn = {2195-0571},
	url = {https://link.springer.com/10.1007/s40295-024-00472-5},
	doi = {10.1007/s40295-024-00472-5},
	language = {en},
	number = {6},
	urldate = {2024-12-18},
	journal = {The Journal of the Astronautical Sciences},
	author = {Michelotti, Niccolò and Rizza, Antonio and Giordano, Carmine and Topputo, Francesco},
	month = dec,
	year = {2024},
	pages = {55},
}

@inproceedings{wolf_multi-fidelity_2022,
	address = {San Diego, CA \& Virtual},
	title = {Multi-{Fidelity} {Uncertainty} {Propagation} for {Objects} in {Cislunar} {Space}},
	isbn = {978-1-62410-631-6},
	url = {https://arc.aiaa.org/doi/10.2514/6.2022-1774},
	doi = {10.2514/6.2022-1774},
	language = {en},
	urldate = {2026-03-17},
	booktitle = {{AIAA} {SCITECH} 2022 {Forum}},
	publisher = {American Institute of Aeronautics and Astronautics},
	author = {Wolf, Trevor and Zucchelli, Enrico M. and Jones, Brandon A.},
	month = jan,
	year = {2022},
}

@article{luo2017review,
  title={A review of uncertainty propagation in orbital mechanics},
  author={Luo, Ya-zhong and Yang, Zhen},
  journal={Progress in Aerospace Sciences},
  volume={89},
  pages={23--39},
  year={2017},
  publisher={Elsevier}
}

@article{isserlis1918formula,
  title={On a formula for the product-moment coefficient of any order of a normal frequency distribution in any number of variables},
  author={Isserlis, Leon},
  journal={Biometrika},
  volume={12},
  number={1/2},
  pages={134--139},
  year={1918},
  publisher={JSTOR}
}

@book{olver2010nist,
  title={NIST handbook of mathematical functions hardback and CD-ROM},
  author={Olver, Frank WJ},
  year={2010},
  publisher={Cambridge university press}
}

@book{scheeres2016orbital,
  title={Orbital motion in strongly perturbed environments: applications to asteroid, comet and planetary satellite orbiters},
  author={Scheeres, Daniel J},
  year={2012},
  publisher={Springer}
}

@article{sudret2008global,
  title={Global sensitivity analysis using polynomial chaos expansions},
  author={Sudret, Bruno},
  journal={Reliability engineering \& system safety},
  volume={93},
  number={7},
  pages={964--979},
  year={2008},
  publisher={Elsevier}
}

@article{lefebvre2021moment,
  title={On moment estimation from polynomial chaos expansion models},
  author={Lefebvre, Tom},
  journal={IEEE Control Systems Letters},
  volume={5},
  number={5},
  pages={1519--1524},
  year={2021},
  publisher={IEEE}
}

@article{cornish1938moments,
  title={Moments and cumulants in the specification of distributions},
  author={Cornish, Edmund A and Fisher, Ronald A},
  journal={Revue de l'Institut international de Statistique},
  pages={307--320},
  year={1938},
  publisher={JSTOR}
}

@article{vittaldev2016spacecraft,
  title={Spacecraft uncertainty propagation using gaussian mixture models and polynomial chaos expansions},
  author={Vittaldev, Vivek and Russell, Ryan P and Linares, Richard},
  journal={Journal of Guidance, Control, and Dynamics},
  volume={39},
  number={12},
  pages={2615--2626},
  year={2016},
  publisher={American Institute of Aeronautics and Astronautics}
}

@inproceedings{adurthi2012conjugate,
  title={The conjugate unscented transform—an approach to evaluate multi-dimensional expectation integrals},
  author={Adurthi, Nagavenkat and Singla, Puneet and Singh, Tarunraj},
  booktitle={2012 American Control Conference (ACC)},
  pages={5556--5561},
  year={2012},
  organization={IEEE}
}

@article{jones2013nonlinear,
  title={Nonlinear propagation of orbit uncertainty using non-intrusive polynomial chaos},
  author={Jones, Brandon A and Doostan, Alireza and Born, George H},
  journal={Journal of Guidance, Control, and Dynamics},
  volume={36},
  number={2},
  pages={430--444},
  year={2013},
  publisher={American Institute of Aeronautics and Astronautics}
}

@article{boone2023directional,
  title={Directional state transition tensors for capturing dominant nonlinear effects in orbital dynamics},
  author={Boone, Spencer and McMahon, Jay},
  journal={Journal of Guidance, Control, and Dynamics},
  volume={46},
  number={3},
  pages={431--442},
  year={2023},
  publisher={American Institute of Aeronautics and Astronautics}
}

@book{vallado_book,
  title={Fundamentals of Astrodynamics and Applications},
  author={David A. Vallado},
  year={2013},
  publisher = {Microcosm Press, Springer},
address = {Hawthorne, CA, USA}
}

@article{ferrariTrajectoryOptions, 
title="{Trajectory options for Hera’s Milani CubeSat Around (65803) Didymos}", 
volume={68}, 
DOI={10.1007/s40295-021-00282-z}, 
number={4}, 
journal={The Journal of the Astronautical Sciences}, 
author={Ferrari, Fabio and Franzese, Vittorio and Pugliatti, Mattia and Giordano, Carmine and Topputo, Francesco}, 
year={2021}, 
pages={973-–994}
}

@article{kelly2024robust,
  title={Robust Cislunar Trajectory Optimization in the Presence of Stochastic Errors},
  author={Kelly, Scott and Geller, David},
  journal={The Journal of the Astronautical Sciences},
  volume={71},
  number={4},
  pages={30},
  year={2024},
  publisher={Springer}
}

\end{document}